\documentclass[preprint,11pt,authoryear]{elsarticle}

\usepackage[utf8]{inputenc}
\usepackage[T1]{fontenc}
\usepackage[english]{babel}
\usepackage{microtype}

\usepackage{amsmath,amssymb}
\usepackage{siunitx}
\usepackage[version=4]{mhchem}   

\usepackage{booktabs}
\usepackage{tabularx}
\usepackage{multirow}
\usepackage{array}

\usepackage{graphicx}
\usepackage{tikz}
\usetikzlibrary{positioning,shapes,shapes.geometric,arrows.meta,fit,backgrounds,calc}

\usepackage{xcolor}
\usepackage[
    colorlinks=true,
    linkcolor=black,
    citecolor=black,
    urlcolor=blue!60!black,
    breaklinks=true
]{hyperref}
\usepackage{cleveref}

\newcommand{\oco}{OCO}
\newcommand{\levelzero}{\textbf{L0}}
\newcommand{\levelone}{\textbf{L1}}
\newcommand{\leveltwo}{\textbf{L2}}
\newcommand{\levelthree}{\textbf{L3}}
\newcommand{\bntbt}{BNT-BT}

\journal{arXiv}

\makeatletter
\def\ps@pprintTitle{%
    \let\@oddhead\@empty
    \let\@evenhead\@empty
    \let\@oddfoot\@empty
    \let\@evenfoot\@empty
}
\makeatother

\begin{document}

\begin{frontmatter}

\title{A Multi-Level Architecture for Reusable Materials Ontologies\\
       --- The OntoCrafter Ceramics Ontology (OCO) as Reference Implementation}

\author[numberland]{Thomas Pannek}

\author[numberland]{Wolfgang Grond\corref{cor1}}
\ead{grond@numberland.com}

\cortext[cor1]{Corresponding author.}

\affiliation[numberland]{
    organization={Numberland --- Dr. Wolfgang Grond},
    addressline={Lohfeld 20},
    postcode={95326},
    city={Kulmbach},
    country={Germany}
}

\begin{abstract}
The Materials Science and Engineering ontology landscape is
fragmented along multiple axes simultaneously. Horizontally: a
recent survey identified 94 ontologies of which over 40 are
structurally incompatible; each new application domain ---
ceramics, polymers, batteries, smart materials --- typically
restarts ontology design from scratch. Vertically: EU regulation
(CSRD, CSDDD, PPWR, CBAM, R2R, AI Act, ESPR) forces material,
manufacturing, supply-chain, and lifecycle data into integrated
digital product passports, leaving ontologies that only address
horizontal fragmentation incomplete for any contemporary
consumer. And mechanistically: a vocabulary that records that
\bntbt{} has $d_{33} \approx 580\,\text{pC/N}$ stores a fact but
cannot surface \emph{why} --- Bi-6s$^2$ lone-pair stereo-activity,
anomalous Born effective charges, soft modes, defect chemistry
--- without a systematic explanation skeleton. We propose a
multi-level modular architecture with two independent classification
axes --- level of abstraction (\levelzero{} bridges, \levelone{}
material-agnostic laboratory-notebook, \leveltwo{}
material-class-specific, \levelthree{} categorical reasoning) and
consumer audience (material vs.\ compliance) --- in which the
material-specific level is internally organised by a seven-tier
mechanistic-explanation skeleton (Symmetry, Energy/DFT,
Thermo/CALPHAD, Kinetics, Microstructure, Defect chemistry,
Bonding) applicable to any crystalline ionic oxide. The
level-and-audience modularity dissolves the horizontal
fragmentation, the compliance audience absorbs the vertical
regulation pressure, and the seven-tier organisation of Level~2
delivers the mechanistic explanation depth. We instantiate the architecture as the OntoCrafter
Ceramics Ontology (\oco{}~v0.94): 5\,196 classes across 44
modules; 167\,348 OWL axioms (40\,454 logical) including 5\,920
reified Neumann tensor constraints; 1\,674 properties; 829
cross-ontology bridge mappings across 40 sections; 1\,172 SHACL
shapes; 163 published competency questions with 52 executable
SPARQL tests (52/52 pass). Functional ceramics are the reference
material domain with \bntbt{} as the end-to-end pilot
demonstrated across all seven explanation tiers; ferritic
high-performance ceramics are in active development as the
second material instance. v0.94 is the release engineered to
enter productive practice; v1.0 is reserved for the state after
that practice has fed back corrections.
\end{abstract}


\end{frontmatter}



\section{Introduction}
\label{sec:introduction}

The design of an industrial materials ontology faces three simultaneous
challenges. None is new in isolation; what is new is that an ontology
entering productive use today must answer all three at once.

The first challenge is the \emph{horizontal fragmentation of the
materials-science ontology landscape}.
\citet{norouzi_2024_landscape} identify 94 ontologies in the field,
of which over 40 are structurally incompatible
\citep{rajamohan_2025_mdsonto}. Within the German Platform
MaterialDigital (PMD) consortium%
\footnote{The PMD Phase-1 publication portfolio sets the practical
context for this work: \citep{bardehle_2025_pmd01,
bekemeier_2025_pmd02, raabe_2022_pmd03, gramlich_2024_pmd04,
chakraborty_2024_pmd05, tomkovic_2026_pmd06, schilling_2026_pmd07,
beyginasrabadi_2025_pmd08, leipner_2025_pmd09, gumbsch_2025_pmd10,
bayerlein_2024_pmd11, bekemeier_2025_pmdworkflows,
beyginasrabadi_2024_pmd13, schilling_2024_pmd14, shoghi_2024_pmd15,
schilling_2024_pmd16, roters_2024_pmd17, fliegener_2024_pmd18,
nerella_2024_pmd19, eisenbart_2024_pmd20, sajjad_2024_pmd21,
beyginasrabadi_2024_pmd22, nahshon_2024_pmd23,
bjarsch_2024_pmd24, mieller_2024_ferrite, mieller_2024_pmd26,
chen_2024_pmd27, luger_2024_pmd28, aschemann_2024_pmd29,
nebel_2024_pmd30, beran_2024_pmd31, maas_2024_smadi,
oezcep_2024_shacl, bayerlein_2021_pmd34, gogula_2024_pmd35,
pan_2023_pmd36, arendt_2024_pmd37, bayerlein_2024_pmd38,
beyginasrabadi_2024_pmd39, beyginasrabadi_2024_pmd40,
bayerlein_2024_pmd41, beyginasrabadi_2024_pmd42,
bayerlein_2024_pmdco, meng_2024_pmd44, agrawal_2024_pmd45,
unger_2023_pmd46, blum_2023_pmd47, beyginasrabadi_2023_pmd48,
schmidtke_2023_pmd49, valdestilhas_2023_pmd50,
mertens_2024_pmd51, chen_2024_pmd52, mutz_2022_pmd53,
garabedian_2024_pmd54, rego_2022_pmd55, lizarazu_2024_pmd56,
rogge_2024_pmd57}.}, more than a dozen projects ---
KnowNow \citep{benhassine_2024_knownow},
Mieller-Ferrit \citep{mieller_2024_ferrite},
SmaDi \citep{maas_2024_smadi},
KupferDigital, GlasDigital, StahlDigital, DiProMag, iBain
\citep{bekemeier_2025_pmdworkflows} ---
each developed its own material-specific ontology, repeatedly
re-modelling laboratory workflows, equipment, and measurement
methods that are essentially identical across all of them. Only a
fraction of the modelled knowledge is material-specific (the actual
class hierarchy of ceramics, polymers, batteries, \ldots); the
largest part --- workflow provenance, equipment, methods,
identifier schemes --- is material-agnostic and reusable.

The second challenge is the \emph{vertical convergence pressure
driven by EU regulation}. The Corporate Sustainability Reporting
Directive (CSRD/ESRS), the Corporate Sustainability Due Diligence
Directive (CSDDD), the Packaging and Packaging Waste Regulation
(PPWR), the Carbon Border Adjustment Mechanism (CBAM), the
Right-to-Repair Directive (R2R), the EU AI Act, the Construction
Products Regulation (CPR), and the Ecodesign for Sustainable
Products Regulation (ESPR) all require digital product passports
(DPPs) that combine material with manufacturing and value-chain
information. A materials ontology that addresses only horizontal
fragmentation (per material class) without this vertical extension is
incomplete for any consumer operating under contemporary EU
regulation. The convergence is visible in the structure of the PMD
programme itself: PMD Phase-3 explicitly incorporates the
\emph{Wertschöpfungskette} (value chain) as a programme
requirement, anticipating integration with Manufacturing-X.

The third challenge is \emph{mechanistic explanation depth}. A
materials ontology that records that \bntbt{} has
$d_{33} \approx 580\,\text{pC/N}$ at its morphotropic phase
boundary stores a fact. An ontology that can also surface
\emph{why} --- Bi-6s$^2$ lone-pair stereo-activity, anomalous Born
effective charges, soft-mode at the Brillouin-zone centre,
R3c$\leftrightarrow$P4mm domain coexistence,
V$_\text{O}^{\bullet\bullet}$ defect chemistry, Hall-Petch
grain-boundary mediation, Mn-acceptor pinning of domain-wall
mobility --- provides authoritativeness that a description-only
vocabulary cannot. The challenge is to attach mechanistic
explanation classes systematically without inflating the
vocabulary into a textbook.

We answer these three challenges with a single integrated design:
a \emph{multi-level, modular ontology with two independent classification
axes plus a seven-tier mechanistic-explanation skeleton internal to
the material-specific level}. The first axis is the \textbf{level of
abstraction} (\levelzero{} bridges, \levelone{} material-agnostic
laboratory-notebook, \leveltwo{} material-class-specific,
\levelthree{} categorical reasoning). The second is the
\textbf{consumer audience} (material versus compliance). The
material-specific level (\leveltwo{}) is internally organised by a
\textbf{seven-tier mechanistic-explanation skeleton} applicable to
any crystalline ionic oxide: Symmetry, Energy/DFT, Thermo/CALPHAD,
Kinetics, Microstructure, Defect chemistry, Bonding. The two axes
are independent --- a module is placed on each at the level
appropriate to its content, and consumers select which subsets they
need --- while the seven-tier skeleton structures the depth of
mechanistic content within Level~2. The level-and-audience
modularity dissolves the horizontal fragmentation, the compliance
audience absorbs the vertical EU-regulation pressure, and the
seven-tier organisation of the material-specific level delivers the
mechanistic explanation depth --- three challenges, three
architecturally distinct answers, without collapsing them onto a
single axis where they would compete.

We instantiate this design as the OntoCrafter Ceramics Ontology
(\oco{} v0.94), the release engineered to enter productive use.
\oco{} ships as a class-bearing module set with bridge mappings,
SHACL shapes, a competency-question catalogue with executable
SPARQL tests, and architecture decision records; the full metric
inventory and method-of-measurement is in
\Cref{sec:implementation}. Functional ceramics are the reference
material domain, with \bntbt{} as the end-to-end pilot demonstrated
across all seven explanation tiers. v0.94 enters productive
practice; corrections surfaced by that practice are the planned
input for v1.0.

This paper contributes:
\begin{enumerate}
\item A formal description of the four-level modular architecture,
      driven by ten engineering design principles (modularity,
      adaptability, interoperability, purpose, equality,
      compatibility, functionality, authoritativeness,
      facetedness, portability), of which five go beyond the
      Norouzi REQ-canon (\Cref{sec:positioning},
      \Cref{sec:architecture}). The same modular pattern addresses
      horizontal material-domain fragmentation and vertical
      EU-regulatory convergence with one structural primitive.
\item The seven-tier mechanistic explanation skeleton
      formalised as the internal organising principle of
      Level~2 (material-specific knowledge), generic for
      crystalline ionic oxides, with cross-tier SHACL
      constraints that make the explanation chain auditable
      (\Cref{sec:7layer}).
\item A reference implementation as the OntoCrafter Ceramics
      Ontology (\oco{}~v0.94), realising the architecture across
      class-bearing material modules; a Neumann tensor engine that
      generates reified symmetry constraints algorithmically over
      the 32 crystallographic point groups; SPARQL-based
      phase-state coupling that derives the active point group
      from sample temperature and composition; a
      coordination-polyhedron module
      (\texttt{oco-localstructure}) that closes the mechanistic
      chain from composition through local geometry to tensor
      symmetry; multi-axis parameter classification along role,
      reference, and material-abstraction layer; an external-cache
      pattern for high-volume reference data (Shannon ionic radii,
      IUCr bond-valence parameters, pyxtal Wyckoff positions,
      Materials Project DFT corpus, Pauling electronegativities);
      and a disciplined reuse-before-invention bridge policy
      (\Cref{sec:implementation}).
\item A vertical extension of the same architecture into the
      compliance and value-chain stack: modules for Life Cycle
      Assessment (with ecoinvent and EN~15804+A2 bridges), CSRD/ESRS
      reporting, supply-chain due diligence (CSDDD), packaging
      (PPWR), carbon-border adjustment (CBAM), right-to-repair
      (R2R), the EU AI Act, ODRL/trust, Manufacturing-X identifier
      and traceability infrastructure (AAS IEC~63278, Catena-X
      CX-0010/CX-0146), and laboratory automation (SiLA). The same
      modular level discipline absorbs these without restructuring
      the material core (\Cref{sec:vertical_extension}).
\item Validation against the Norouzi quality-requirement canon
      (REQ1--REQ9, all nine met), the OOPS! pitfall audit
      \citep{oops_pitfall_scanner}, a published catalogue of
      area-tagged competency questions with executable SPARQL
      tests, and a SHACL validator suite (counts in
      \Cref{sec:implementation}; reasoner/audit results in
      \Cref{sec:validation}).
\item Discussion of limitations and the v1.0 roadmap, framed by
      the corrections expected from productive practice ---
      including a second ceramic material system (ferritic
      high-performance ceramics) currently in active development
      (\Cref{sec:discussion}).
\end{enumerate}

The remainder of the paper is organised as follows.
\Cref{sec:related_work} surveys the MSE ontology landscape,
positions the multi-level architecture against three established
schools of design (BFO-aligned mid-level, EMMO-centric, and
bottom-up domain-specific), and lays out the ten design principles
that drive \oco{} (\Cref{sec:positioning}). \Cref{sec:architecture}
presents the four-level architecture together with the
material/compliance audience axis as the second classification
dimension, plus the seven-tier mechanistic-explanation skeleton
as the internal organisation of Level~2. \Cref{sec:implementation} describes \oco{} as
reference implementation --- the material modules (including the
Neumann engine, phase-state coupling, multi-axis parameter
classification, the coordination-polyhedron module, and the
seven-tier explanation skeleton with its external-cache pattern),
the bridge inventory, and the vertical extension into the
compliance and value-chain stack. \Cref{sec:validation} reports the
validation results --- all nine Norouzi requirements met.
\Cref{sec:discussion} discusses strengths, limitations, and the
v1.0 horizon shaped by productive practice. \Cref{sec:conclusion}
summarises and outlines the roadmap.


\section{Background and Related Work}
\label{sec:related_work}

\subsection{The State of MSE Ontologies}

\citet{norouzi_2024_landscape} provide the most comprehensive recent
survey of MSE ontologies, evaluating 94 semantic artifacts (4
top-level, 8 mid-level, 60 domain-level, 2 application-level, plus
20 not openly evaluable) against nine quality requirements (REQ1--9)
and an OOPS! pitfall audit. Their key findings frame the work
presented here:

\begin{itemize}
\item Only 9 of 94 ontologies (10\,\%) publish competency questions.
\item No ontology adopts user stories or personas as design anchors.
\item Critical OOPS! pitfalls (P19 multiple-domain/range, P40
      namespace-hijacking, P31 incorrect \texttt{equivalentClass},
      P11 missing domain/range) are widespread, even in established
      ontologies such as QUDT (217 P11 hits).
\item BFO is the most-reused top-level (16$\times$); EMMO follows
      (12$\times$); PMDco is reused only $1\times$ directly despite
      being deployed in 13 BMBF projects.
\end{itemize}

\subsection{Three Architectural Schools}

\noindent\textbf{Schule A --- BFO-aligned Mid-Level.} PMDco
\citep{bayerlein_2024_pmdco}, MWO \citep{nfdi_matwerk_mwo}, MSEO,
NFDIcore. Strict ISO/IEC 21838-2 conformance, mid-level
material-science vocabulary above BFO, formal reasoner-friendly.
Cost: high BFO learning curve, ``portion-of'' constructs awkward
for non-philosophers.

\noindent\textbf{Schule B --- EMMO-centric.} BattINFO,
CHAMEO \citep{delnostro_2022_chameo}, GPO, EMMO Crystallography,
\texttt{domain-electrochemistry}. Adopted in EU projects (EMMC,
BIG-MAP, OntoTrans). Cost: each subdomain imports the full EMMO
top-level; many subdomains are at v0.x maturity.

\noindent\textbf{Schule C --- Bottom-up domain-specific.}
KnowNow \citep{benhassine_2024_knownow}, MaterialDigital project
ontologies, NanoMine, MatKG. Pragmatic and fast, but inter-project
interoperability weak.

\subsection{Specific Gaps in the Current Landscape}

Beyond the methodological pitfalls flagged by
\citet{norouzi_2024_landscape}, two content gaps emerge from a
systematic reading of the recent literature:

\noindent\textbf{No deep functional-ceramics modeling.} KnowNow is
restricted to LTCC multi-layer components; Mieller et al. cover only
NiCuZn ferrite \citep{mieller_2024_ferrite}; BattINFO addresses
batteries; SmaDi \citep{maas_2024_smadi} models four smart-material
classes but at the level of constitutive parameters, not microstructure
or defect chemistry. The combination of 230 space groups,
Kröger-Vink defect notation \citep{kroeger_vink_1956}, Newnham
composite connectivity \citep{newnham_1978_connectivity}, and a full
coupled-effects family (piezoelectric, pyroelectric, magnetostrictive,
\ldots) is absent from all surveyed ontologies.

\noindent\textbf{Integrity constraints not delivered alongside the
TBox.} OWL's open-world semantics interprets missing triples as
unknown rather than as violations --- a deliberate and correct
design choice for an open-world reasoning language, but one that
leaves enforceable integrity constraints (mandatory fields, value
ranges, cardinality caps) outside what TBox axioms can express.
The standard remedy is SHACL, which adds closed-world shape
validation as an independent artifact type alongside the TBox;
\citet{oezcep_2024_shacl} demonstrate the pattern with their
SHACL-OBDA validator. The gap in the current MSE ontology
landscape is not the Open-World Assumption itself --- it is the
correct semantics for what OWL is for --- but that most surveyed
MSE ontologies ship a TBox without accompanying SHACL shapes,
leaving consumers without a standard way to validate ABox
compliance.

\subsection{Positioning of OCO}
\label{sec:positioning}

\oco{}'s architecture is governed by ten design principles drawn
from engineering practice (platform strategy, construction
catalogues), quality management (SIPOC), and the materials-science
domain itself. These principles, not the Norouzi requirement canon,
are the \emph{primary} design drivers; the canon is a verification
surface that the principles must clear, not a brief from which they
derive. Two emphases run through the principle set: a
\emph{lessons-learned} dimension that distills what worked and what
did not in the PMD Phase-1 ontology cohort
(\Cref{sec:related_work}), and a \emph{forward-looking} dimension
that anticipates the EU-driven convergence of MaterialDigital with
Manufacturing-X --- the second axis of fragmentation introduced in
\Cref{sec:introduction}. We list the principles here because they
explain the architectural choices that follow in
\Cref{sec:architecture}, and because five of them go beyond what
REQ1--9 demand.

\subsubsection{Ten Design Principles}

\begin{figure}[htbp]
  \centering
  \resizebox{\linewidth}{!}{

\begin{tikzpicture}[
    every node/.style={font=\footnotesize},
    phase/.style={
        rectangle, rounded corners=2pt,
        draw=black!60, line width=0.6pt, fill=blue!8,
        minimum height=2.0cm, minimum width=3.6cm,
        text width=3.4cm, align=center, inner sep=4pt
    },
    principles/.style={
        rectangle, rounded corners=2pt,
        draw=black!60, line width=0.4pt, fill=yellow!15,
        minimum width=3.6cm, text width=3.4cm,
        align=center, font=\scriptsize, inner sep=4pt
    },
    arrow/.style={
        -{Latex[length=2.5mm,width=2mm]},
        line width=1pt, draw=blue!55!black
    },
    timeline/.style={font=\scriptsize\itshape\color{black!55}}
  ]

  \node[timeline] at (0, 2.55) {Consumer needs};
  \node[timeline] at (12, 2.55) {v0.94 $\to$ v1.0+ extension};
  \draw[black!25, dashed, very thin] (-1.9, 2.25) -- (13.9, 2.25);

  \node[phase] (p1) at (0, 0.9) {%
    \textbf{Modular consumer choice}\\[3pt]
    Selective import depth; reasoning cost matches the level loaded
  };
  \node[phase] (p2) at (4, 0.9) {%
    \textbf{External anchoring}\\[3pt]
    Bridge to ELN/LIMS, ML datasets, PROV-O, Manufacturing-X
  };
  \node[phase] (p3) at (8, 0.9) {%
    \textbf{Authoritative domain depth}\\[3pt]
    Explain \emph{why} a property holds, not only \emph{that} it does
  };
  \node[phase] (p4) at (12, 0.9) {%
    \textbf{Sister-domain extension}\\[3pt]
    \levelone{}/\leveltwo{} replaceability across material classes
  };

  \draw[arrow] (p1.east) -- (p2.west);
  \draw[arrow] (p2.east) -- (p3.west);
  \draw[arrow] (p3.east) -- (p4.west);

  \node[principles] (pr1) at (0, -1.7) {%
    (1)~Modularity\\
    (7)~Functionality
  };
  \node[principles] (pr2) at (4, -1.7) {%
    (2)~Adaptability\\
    (3)~Interoperability
  };
  \node[principles] (pr3) at (8, -1.7) {%
    (4)~Purpose\\
    (5)~Equality\\
    (8)~Authoritativeness\\
    (9)~Facetedness
  };
  \node[principles] (pr4) at (12, -1.7) {%
    (6)~Compatibility\\
    (10)~Portability
  };

  \draw[black!40, dashed, line width=0.5pt] (p1.south) -- (pr1.north);
  \draw[black!40, dashed, line width=0.5pt] (p2.south) -- (pr2.north);
  \draw[black!40, dashed, line width=0.5pt] (p3.south) -- (pr3.north);
  \draw[black!40, dashed, line width=0.5pt] (p4.south) -- (pr4.north);

  \node[font=\scriptsize\itshape\color{black!55}] at (6, -3.2)
    {Principles that specifically enable each phase};

\end{tikzpicture}}
  \caption{Causal-chain view of OCO's ten design principles. The
  four phases (top row) trace the consumer-need-to-extension flow
  that the architecture must answer; each phase is enabled by a
  specific cluster of principles (bottom row). Numbers refer to
  the principle list below.}
  \label{fig:principles_story}
\end{figure}
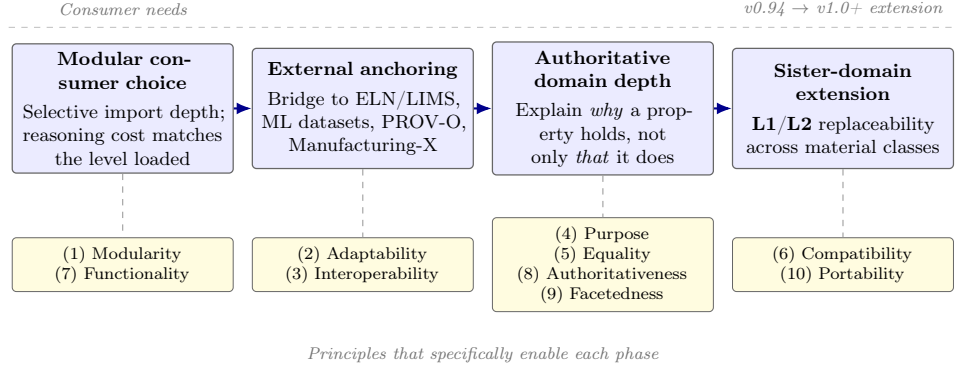

\Cref{fig:principles_story} organises the ten principles below as
a causal chain: each principle cluster enables the next phase of
what an industrial materials ontology must deliver, from
consumer-side modular choice all the way through to sister-domain
extension. The detailed list that follows is grouped in the same
order.

\begin{enumerate}
\item \textbf{Modularity.} A materials ontology should follow the
      same logic as engineering platform strategy: classes and
      properties are organized into reusable modules that can be
      combined per consumer, analogous to the construction catalogues
      used by automotive platforms. A consumer who needs only
      equipment and method vocabulary imports two modules, not the
      whole ontology.

\item \textbf{Adaptability.} Bridges to external ontologies are
      isolated in a dedicated level (\levelzero{}), and the imported
      external sources are versioned and cached locally. When an
      external ontology releases a new version, only the bridge file
      changes --- no class definitions in \oco{} break, and the
      cache makes the build reproducible. The same mechanism extends
      \emph{forward}: standards that are still consolidating (EU
      Construction Products Regulation 2024/3110, Critical Raw
      Materials Act, evolving DPP-XBRL schemas) can be tracked via
      provisional bridges that mature alongside the upstream
      release, without forcing rework in the material core.

\item \textbf{Interoperability.} An ontology is not a free-standing
      artifact but a vocabulary embedded in a workflow. For a
      materials ontology this requires concrete anchoring in the
      open-source materials stack: ELN/LIMS, machine-learning datasets
      (Croissant), materials databases (OPTIMADE), and provenance
      tooling (PROV-O).

\item \textbf{Purpose.} Vocabularies without a stated purpose
      accumulate dead classes. \oco{} binds its scope to a published
      catalogue of competency questions, each tagged with an
      \emph{area} (tensor symmetry, phase state, route, lifecycle,
      \ldots) and accompanied by an executable SPARQL test where
      possible.

\item \textbf{Equality.} Experimental documentation is never
      complete; any record is an SIPOC fragment --- who was the
      supplier of the inputs, what was the process, what was the
      output, who is the customer. Scientific publications are
      treated identically: a paper is an SIPOC fragment of one or
      more experiments. The same provenance pattern applies to both,
      which removes the artificial line between raw experimental data
      and published results.

\item \textbf{Compatibility.} \oco{} is the architectural successor
      of two earlier PMD-Phase-1 ontologies (KnowNow for LTCC
      multi-layer ceramics, SmaDi for smart materials including
      piezo-ceramic) and the EU-regulatory peer of the
      Manufacturing-X ecosystem (AAS asset administration shells,
      Catena-X BPN identifiers, ESPR-driven sector DPPs). Backward
      compatibility --- ingesting Phase-1 ontologies via bridges
      rather than re-modelling --- and forward compatibility ---
      anchoring in Manufacturing-X identifier and traceability
      infrastructure via bridges to the same kind of pattern ---
      are the same requirement under the same kit
      (\Cref{sec:vertical_extension}).

\item \textbf{Functionality.} Beyond reusable modules, \oco{} is
      organized into four \emph{functional layers} --- \levelzero{}
      (bridges), \levelone{} (laboratory-notebook level,
      material-agnostic, ELN/LIMS-ready), \leveltwo{}
      (material-specific knowledge), \levelthree{} (reasoning axioms)
      --- each addressing a different consumer concern. Maintenance
      becomes localized: a PMDco update touches only \levelzero{}; a
      new material adds only \leveltwo{}.

\item \textbf{Authoritativeness.} For functional ceramics the
      \emph{full mechanistic-explanation chain} --- from local
      structure and bonding through symmetry, energy landscape,
      thermodynamic phase regions, kinetics, microstructure, and
      defect chemistry to the resulting tensor-component symmetry
      --- is load-bearing knowledge. The seven-tier skeleton
      (\Cref{sec:7layer}) is the systematic carrier of that chain.
      A central traversal through it is \emph{phase diagram $\to$
      active crystallographic phase $\to$ tensor-component
      symmetry}: without that traversal an ontology cannot answer
      ``which piezoelectric coefficients are non-zero for \bntbt{}
      at room temperature near the morphotropic phase boundary?''
      But this is one path of several --- defect chemistry to
      Mn-acceptor pinning of domain-wall mobility, soft modes at the
      Brillouin-zone centre to incipient ferroelectricity,
      microstructure to Hall-Petch toughening --- and a vocabulary
      that cannot deliver any of them along its central domain
      questions is not authoritative. \oco{} models the entire
      seven-tier chain explicitly, with the Neumann tensor engine,
      phase regions, and point-group bindings carrying the
      symmetry-to-tensor edge as one example implementation. The
      chain is not ceramics-specific: Neumann's principle and the
      layered explanation pattern hold for every crystalline
      material, so the same mechanism governs metals (martensitic
      transformations in shape-memory alloys, soft-magnetic alloys
      across their Curie transition) and crystalline organics
      (pharmaceutical polymorphs, organic semiconductors). A
      sister-domain \leveltwo{} replacement (\Cref{sec:reuse_wip})
      inherits this entire infrastructure unchanged --- only the
      concrete phase-region populations and the domain-specific
      seven-tier content differ.

\item \textbf{Facetedness.} A material parameter is not adequately
      described by a single classification axis. \oco{} classifies
      every parameter along three independent axes simultaneously:
      \textbf{role} (state, response, transport, structure,
      statistical, topological, fit), \textbf{reference}
      (fundamental, material, interface, defect, measurement,
      specimen, process, environment), and \textbf{layer} (atomic,
      crystalline, microstructural, mesoscopic, macroscopic). The
      axes are populated by reasoner-classified
      \texttt{rdfs:subClassOf} relations, so SPARQL queries can
      navigate any subset of facets without bespoke join logic.

\item \textbf{Portability.} The architecture must support replacement
      of the material-specific level without touching \levelzero{},
      \levelone{}, or \levelthree{}. A metallurgy ontology adopting
      the same \levelzero{}/\levelone{}/\levelthree{} need only
      contribute its own \leveltwo{} module to deliver a fully
      reasoner-ready ontology for its domain.
\end{enumerate}

\subsubsection{Where OCO Goes Beyond the Landscape Canon}

The Norouzi REQ1--9 canon emphasizes reasoning, modularity, bridges,
and documentation --- necessary but not sufficient for a working
materials ontology. Five of the ten principles above are \emph{not}
captured by the canon and reflect requirements that surfaced from
real consumer use cases rather than from the survey methodology:

\begin{itemize}
\item \textbf{Adaptability} is not explicitly in REQ1--9. The canon
      recognizes the importance of reuse but treats external version
      drift as a documentation matter. \oco{}'s strategy
      (bridge-isolation plus external version cache) is an engineering
      response to the empirical observation that bridge files are the
      \emph{only} component routinely broken by external releases.

\item \textbf{Equality} is absent from the canon. The canon
      implicitly distinguishes experimental data from literature;
      \oco{} unifies both as \texttt{prov:Activity} instances with
      the same provenance pattern. This is the architectural
      prerequisite for the SIPOC-Extractor companion tool and for
      cross-paper aggregation without forcing premature reconciliation
      of conflicting reports.

\item \textbf{Compatibility} with predecessor projects (KnowNow,
      SmaDi) addresses a fragmentation problem the canon identifies
      but does not solve: even within a single consortium, every
      project re-models the same laboratory concepts. \oco{}'s
      bridge level ingests both predecessors verbatim, allowing their
      concepts to coexist with \oco{}'s deeper modeling.

\item \textbf{Authoritativeness} in the functional-ceramics sense
      (the full seven-tier mechanistic-explanation chain, of which
      phase $\to$ tensor coupling is one central traversal) is a
      domain requirement that the canon --- being domain-neutral
      --- cannot generate. Without it, a ceramics ontology cannot
      stand behind its own competency questions about coupled
      effects.

\item \textbf{Facetedness} generalizes what the canon calls ``rich
      semantics'' into a concrete multi-axis classification with
      reasoner-supported navigation. The closest the canon comes is
      requiring ``more than \texttt{rdfs:label}'' --- a much weaker
      bar.
\end{itemize}

The remaining five principles (modularity, interoperability, purpose,
functionality, portability) align with REQ1--9 but are realized with
engineering rather than compliance intent: modularity is platform
strategy, not a list of files; functionality is level separation by
consumer concern, not just file count; purpose is bound to executable
tests, not free-text statements.

\subsubsection{Architectural Consequence}

The four-level architecture that the remainder of this paper
describes is the synthesis of these ten principles, not a direct
answer to REQ1--9. Modularity, functionality, and portability dictate
the multi-level structure itself; adaptability dictates the bridge-level
isolation; interoperability and compatibility dictate the bridge
inventory; purpose dictates the area-tagged competency catalogue;
equality dictates the unified \texttt{prov:Activity} treatment of
experiments and publications; facetedness and authoritativeness
dictate the depth of the material-specific level. The Norouzi canon
is satisfied by this architecture (\Cref{sec:validation}), but as a
verification rather than a brief --- and the architecture extends to
requirements the canon does not raise. The reference implementation
for ceramics demonstrates that the pattern is viable; the modular
structure invites sister-domain implementations (metallurgy,
polymers, batteries) that would share \levelzero{} and \levelone{}
with \oco{} while replacing \leveltwo{}.


\section{The Multi-Level Architecture}
\label{sec:architecture}

The architecture introduced in \Cref{sec:introduction} comprises
two independent classification axes plus a mechanistic-explanation
skeleton internal to the material-specific level. The first and
primary axis is the \textbf{level of abstraction} ---
\levelzero{} bridges, \levelone{} material-agnostic
laboratory-notebook level, \leveltwo{} material-class-specific
knowledge, \levelthree{} categorical reasoning --- governed by the
architectural principles in \Cref{sec:arch_principles} and
detailed level-by-level in
\Cref{sec:level0,sec:level1,sec:level2,sec:level3}. The second
axis is the \textbf{consumer audience}
(\Cref{sec:audience_axis}), which separates the material core from
the compliance and value-chain modules without restructuring
either. The two axes are independent: every module is placed on
each at the level appropriate to its content, and consumers select
the subsets they need. Within \leveltwo{} (the material side), the
content is organised internally by the \textbf{seven-tier
mechanistic-explanation skeleton} (\Cref{sec:7layer}), which lets
the ontology surface \emph{why} a property holds, not only that it
does. Compliance modules have no analogous internal structure ---
they are regulatory rather than mechanistic --- which is why the
seven-tier skeleton is the internal organisation of one specific
level rather than a third independent axis.

\begin{figure}[htbp]
  \centering
  \resizebox{\linewidth}{!}{

\begin{tikzpicture}[
    layer/.style={
        draw, thick, rounded corners=2pt,
        minimum width=8.5cm, minimum height=1.0cm,
        font=\small, align=center
    },
    boundary/.style={
        draw, dashed, rounded corners=2pt,
        minimum width=8.5cm, minimum height=0.65cm,
        font=\footnotesize\itshape, align=center,
        text=gray
    },
    arrow/.style={-{Latex[length=2mm]}, thick, gray},
    annot/.style={font=\scriptsize, gray, align=left}
]

\node[boundary] (l4) at (0, 4.5) {%
    Out of scope --- quantitative mathematics (external compute layer)};

\node[layer, fill=orange!15] (l3) at (0, 3.0) {%
    \textbf{L3 Reasoning} --- categorical axioms,
    route templates, CQ-bound SPARQL
};

\node[layer, fill=red!15] (l2) at (0, 1.5) {%
    \textbf{L2 Material-Class-Specific} --- ceramics:
    material, defect, composite, crystal, phase, element\\[1pt]
    \footnotesize BNT-BT + ferritic ceramics;
    Kröger-Vink defects; 230 space groups;
    Newnham connectivity
};

\node[layer, fill=blue!15] (l1) at (0, 0.0) {%
    \textbf{L1 LIMS/ELN-Ready} --- material-agnostic
    workflow, sample, equipment, measurement, identifier\\[1pt]
    \footnotesize with PROV-O, DCL, GUM, ISA pattern
};

\node[layer, fill=green!15] (l0) at (0, -1.5) {%
    \textbf{L0 Bridge} --- to PMDco, QUDT, PROV-O, CHMO,
    NFDIcore, FaBiO, IAO, OCE, Wikidata, \ldots
};

\draw[arrow] (l0.north) -- (l1.south);
\draw[arrow] (l1.north) -- (l2.south);
\draw[arrow] (l2.north) -- (l3.south);

\node[annot, anchor=west] at (4.5, 3.0) {OWL\,2\,DL};
\node[annot, anchor=west] at (4.5, 1.5) {OWL\,2\,EL};
\node[annot, anchor=west] at (4.5, 0.0) {RDFS};
\node[annot, anchor=west] at (4.5, -1.5) {RDFS};

\end{tikzpicture}}
  \caption{The four-level architecture for reusable materials
           ontologies. Consumers select their import depth: \levelzero{}
           alone for bridge-only integration, \levelzero{}+\levelone{}
           for material-agnostic LIMS/ELN, \levelzero{}+\levelone{}+\leveltwo{}
           for full ceramics knowledge, and the opt-in
           \levelzero{}+\levelone{}+\leveltwo{}+\levelthree{} bundle
           for OWL-DL reasoning. Quantitative mathematics is
           explicitly out of scope for this release
           (\Cref{sec:scope_boundary}).}
  \label{fig:level_stack}
\end{figure}
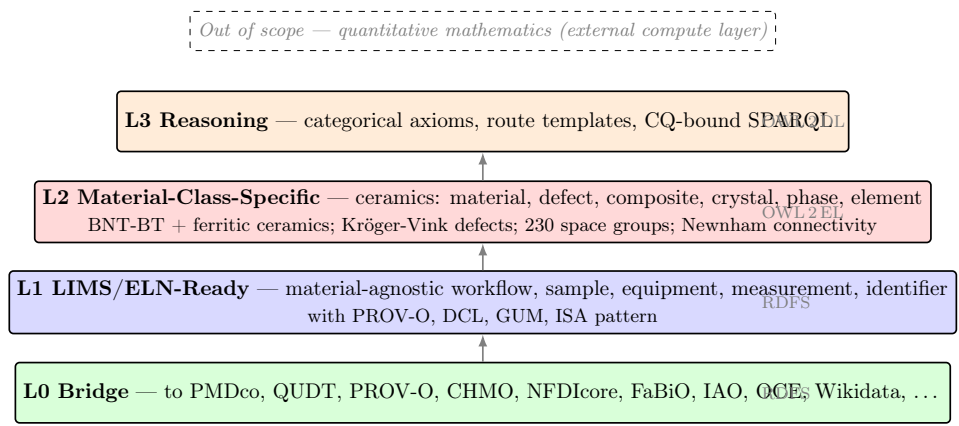

\subsection{Architectural Principles}
\label{sec:arch_principles}

The architecture rests on five design principles:

\begin{enumerate}
\item \textbf{Level separation.} Each level has a clearly bounded
      scope: \levelzero{} bridges to external standards, \levelone{}
      carries the material-agnostic vocabulary, \leveltwo{} the
      material-specific vocabulary, \levelthree{} the categorical
      reasoning. Consumers select their import depth.
\item \textbf{Modularity within levels.} Each level decomposes into
      independently maintainable modules. Consumers can import single
      modules (e.g., \texttt{oco-equipment} alone) without pulling
      the entire level.
\item \textbf{Opt-in reasoning.} Categorical reasoning (\levelthree{})
      is not part of the default distribution. Consumers without
      reasoning needs do not pay the OWL-DL reasoner cost.
\item \textbf{Bridge localization.} External version changes (PMDco
      v3.1, QUDT v3.3) are confined to a single level (\levelzero{}).
      Internal modules reference only \texttt{bridge:*} classes,
      which remain stable under our control.
\item \textbf{Cache pattern for external reference data.} Large
      externally-curated bodies of reference data --- mid-level
      ontology versions consumed by \levelzero{} bridges, structural
      and thermodynamic constants consumed by the mechanistic tiers
      of \leveltwo{} --- are not embedded in OCO's TBox but cached
      locally as versioned snapshots, pinned via SHA in
      \texttt{bridge/external\_versions.yaml}. This separates
      reference-data scale from ontology scale: the TBox stays
      compact, while consumers query large reference corpora
      (Materials Project DFT entries, IUCr bond-valence parameters,
      pyxtal Wyckoff positions, Shannon ionic radii, Pauling
      electronegativities) with reproducible version pinning. Concrete instances appear in
      \Cref{sec:level0} (external ontologies) and \Cref{sec:7layer}
      (mechanistic-tier reference data).
\end{enumerate}

These five principles separate the architecture concretely from
the two dominant schools introduced in \Cref{sec:related_work}.
PMDco offers five distributions
(\textit{full}, \textit{base}, \textit{simple}, \textit{minimal},
\textit{main}), but these are vertical slices through the same
mixed vocabulary --- modular by class count, not by architectural
level. The \levelone{}/\leveltwo{} split that isolates
``laboratory workflow without material classes'' from ``material
description without workflow classes'' along the consumer-concern
axis cannot be expressed in PMDco's distribution scheme. EMMO
sub-domain ontologies (BattINFO, CHAMEO, GPO, the EMMO
crystallography and electrochemistry modules) inherit the full
EMMO top-level (700+ classes) and pay the corresponding reasoner
cost regardless of which classes a given consumer actually uses;
the opt-in reasoning principle realized here via a separate
\levelthree{} distribution cannot be retrofitted onto an EMMO
sub-domain without breaking its EMMO inheritance. The contrast is
therefore not over rival mid-level vocabularies but over the
modularity \emph{axis} (class count versus consumer concern) and
the reasoning \emph{posture} (mandatory inheritance versus opt-in
distribution).

\subsection{Level 0 --- Bridge to External Standards}
\label{sec:level0}

\levelzero{} is the level at which OCO's interoperability claim is
made concrete. Its content is a curated set of \texttt{bridge:*}
classes that act as named anchors between OCO's internal vocabulary
and external mid-level ontologies. Internal modules reference only
these anchors, never the external IRIs directly --- this isolates
external-version churn (a PMDco~v3.1 release, a QUDT update) to a
single level that is small enough to audit class-by-class.

The selection of bridge targets is not opportunistic: a target is
admitted only if it satisfies all of the following criteria.

\begin{enumerate}
\item \textbf{OWL/RDF format.} The target must be available as a
      machine-readable ontology in a standard W3C format. Industrial
      classification schemes published only as proprietary XML or
      catalog formats are excluded.
\item \textbf{Mid-level granularity.} Targets must occupy the
      \emph{mid-level} between abstract foundations and instance
      data. Direct alignment to upper-level foundations is achieved
      transitively via the mid-level targets that already commit to
      them (e.g., BFO via PMDco), avoiding redundant philosophical
      commitments at the OCO surface.
\item \textbf{Established adoption.} The target must be in active
      use within the materials-science or research-data-management
      community, with a versioned release history.
\item \textbf{Bounded import cost.} A bridge entry adds at most a
      handful of anchor classes to OCO. Targets whose useful
      adoption requires importing hundreds of unrelated upper
      classes are inadmissible by this criterion --- it would defeat
      the modular multi-level design.
\item \textbf{License compatibility.} Targets must permit
      redistribution under terms compatible with OCO's
      CC-BY/CC-BY-SA scheme.
\end{enumerate}

\Cref{tab:l0_bridges} lists the eleven bridge targets with
substantial class-level coverage (each anchored to $\geq$14 OCO
classes). A further six targets (IAO, m4i, SOSA, DCAT, W3C-Org,
IUCr-Wiki) are present as minimal anchors (1--2 classes each), and
the explicit cross-ontology mapping inventory in
\texttt{bridge\_mappings.yaml} contributes 467 further mappings,
including bridges to the predecessor PMD ontologies KnowNow and
SmaDi, to OPTIMADE, Croissant and the ELN-Filetype standard, and to
the EMMO sub-modules ISQ, chemistry, and materials
(\Cref{sec:interop_extended}).

\begin{table}[htbp]
  \centering
  \caption{\levelzero{} bridge targets with substantial coverage.
  \emph{Anchors} counts the distinct OCO classes that map to the
  target via \texttt{owl:equivalentClass}, \texttt{rdfs:subClassOf},
  or \texttt{skos:exactMatch}/\texttt{closeMatch}, measured on the
  merged distribution \texttt{oco\_merged.ttl} (2026-05-25). Sorted
  by anchor count.}
  \label{tab:l0_bridges}
  \footnotesize
  \begin{tabularx}{\linewidth}{@{}>{\raggedright\arraybackslash}p{4.0cm}Xr@{}}
    \toprule
    Target & Role in OCO & Anchors \\
    \midrule
    ChEBI \citep{hastings_2016_chebi}
      & chemical entities (compounds, salts, reagents)                & 163 \\
    CHMO \citep{chmo_ontology}
      & characterization methods (XRD, SEM, EIS, \ldots)              & 151 \\
    QUDT QuantityKinds \citep{qudt_2018}
      & physical quantities (kinds and units)                         & 141 \\
    Wikidata Q-IDs \citep{vrandecic_2014_wikidata}
      & universal entity anchor (eponyms, elements, materials)        & 136 \\
    PMDco v3.0.0 \citep{bayerlein_2024_pmdco}
      & materials mid-level (process, sample, property, equipment)    & 105 \\
    PROV-O \citep{prov_o_2013}
      & provenance of activities, agents, entities                    & 68 \\
    NFDIcore \citep{nfdicore_ontology}
      & NFDI cross-domain RDM concepts                                & 32 \\
    DOI namespace \citep{iso_26324_doi}
      & citation infrastructure (literature, datasets)                & 22 \\
    OBI
      & biomedical-investigations vocabulary (assays, instruments)    & 20 \\
    schema.org
      & web vocabulary (organizations, products, identifiers)         & 15 \\
    FaBiO \citep{peroni_2012_fabio_cito}
      & bibliographic-resource typing (Article, Patent, Standard)     & 14 \\
    \midrule
    \textbf{Total} (substantial only)
      & many OCO classes carry anchors to more than one target, so
        the column sums to more than the count of distinct bridged
        OCO classes
      & \textbf{867} \\
    \bottomrule
  \end{tabularx}
\end{table}

\paragraph{Targets explicitly not bridged, and why}
A second list of candidate targets has been screened and rejected.
The rejection categories are themselves part of the architectural
argument.

\begin{itemize}
\item \emph{Foundational ontologies} (BFO~\citep{bfo_2021},
      UFO-A~\citep{guizzardi_2005_ufo}). OCO is BFO-aligned
      transitively via PMDco; re-asserting BFO at the OCO surface
      would duplicate that commitment without adding classes.
      UFO-A's rigid Kind/Role/Phase distinctions would force the
      reclassification of all existing OCO classes into a
      meta-typology that the materials community has not adopted ---
      a change with high cost and no demonstrated benefit for our
      competency questions.
\item \emph{Centralist mid-level competitors}
      (EMMO~\citep{ghedini_2019_emmo} and its sub-domain ontologies
      such as the EMMO quantity, electrochemistry, and atomistic
      modules). EMMO sub-domain ontologies require importing the
      full EMMO core (700+ classes), which contradicts the
      bounded-import-cost criterion above and forecloses the opt-in
      reasoning principle: every consumer pays the full EMMO load
      regardless of need. We discuss the centralist-versus-multi-level
      contrast further in \Cref{sec:discussion}.
\item \emph{Mid-level MSE competitors to PMDco}
      (Fraunhofer MSEO). PMDco~v3 has emerged as the consortial
      convergence target for the BFO-aligned mid-level MSE space,
      with active maintenance and broad German-ecosystem adoption.
      Maintaining a second mid-level MSE anchor in parallel would
      reintroduce the dual-anchor ambiguity in the very level
      whose purpose is to provide a unique external anchor per
      concept; the concepts MSEO covers are reachable via PMDco.
\item \emph{Quantity/unit competitors to QUDT}
      (OM~\citep{rijgersberg_2013_om}, EMMO \texttt{domain-quantity}).
      PMDco has standardized on QUDT as its unit anchor. Maintaining
      a second quantity bridge would introduce ambiguity in the very
      level whose purpose is to provide a unique external anchor per
      concept.
\item \emph{Non-OWL industrial classification schemes}
      (ETIM~\citep{etim_classification},
      eCl@ss~\citep{eclass_classification},
      ISO~15926-14~\citep{iso_15926_14},
      NAMUR~NE~100~\citep{namur_ne100}). These are widely adopted in
      industry but are distributed only as BMEcat-XML, proprietary
      catalogs, or non-OWL formats. No semantic mapping path exists
      that does not first require the external community to publish
      an OWL representation.
\item \emph{ELN/LIMS tool ontologies} (Chemotion, eLabFTW, openBIS,
      Kadi4Mat, NOMAD-Oasis, \ldots). These systems are
      \emph{consumers} of OCO's \levelone{} schema, not peer
      ontologies: their concepts (user sessions, folder permissions,
      signature workflows) are deliberately outside OCO's modeling
      scope.
\end{itemize}

\paragraph{One known gap in the present roster}
\textbf{MeSH} (Medical Subject Headings, U.S.\ National Library of
Medicine) is published in SKOS/RDF and qualifies under our
criteria; it is not bridged because the pharmaceutical sister
domain is not currently a consumer, but no architectural obstacle
exists.

The discipline of writing down both the inclusion criteria and the
explicit rejections --- and of naming the known gaps --- turns
\levelzero{} into the documented contract between OCO and the
surrounding ontology landscape, rather than a loose collection of
opportunistic mappings.

\subsection{Level 1 --- Material-Agnostic LIMS/ELN Level}
\label{sec:level1}

\levelone{} contains all concepts that are independent of the
specific material under investigation: a spray dryer dries
pharmaceutical suspensions as well as ceramic slurries; an XRD
measures metals as well as ceramics; a poling station works on
PVDF polymers as well as on \bntbt{}. The level decomposes into
fourteen pure-\levelone{} modules plus the \levelone{} fragments of
two modules whose classes span \levelone{} and \leveltwo{} via the
multi-axis classification (\Cref{sec:multiaxis}). Consumers import
only the parts they need --- an asset-management tool may load
\texttt{oco-equipment} alone; an LIMS may bundle
\texttt{oco-investigation}, \texttt{oco-process},
\texttt{oco-equipment} and \texttt{oco-measurement} and skip the
rest.
\Cref{tab:l1_modules} lists the modules and the architectural reason
for each.

\begin{table}[htbp]
  \centering
  \caption{The \levelone{} modules (all under the \texttt{oco-*}
  namespace). \emph{Classes} counts the \levelone{}-level classes
  per module in the merged distribution; modules marked $\dagger$
  also carry \leveltwo{} classes (concrete materials, tensor-component
  sub-properties --- their \leveltwo{} portion is reported in
  \Cref{tab:l2_modules}).}
  \label{tab:l1_modules}
  \scriptsize
  \begin{tabularx}{\linewidth}{@{}>{\raggedright\arraybackslash}p{1.9cm}>{\raggedleft\arraybackslash}p{0.8cm}>{\raggedright\arraybackslash}X@{}}
    \toprule
    Module & Classes & Why this module is in \levelone{} \\
    \midrule
    \texttt{measurement}
      & 554
      & Characterization-method taxonomy (XRD, SEM, EIS, $d_{33}$,
        impedance spectroscopy, \ldots); methods are
        material-agnostic, what differs is the sample. \\
    \texttt{property}$\dagger$
      & 429
      & Physical-quantity backbone for laboratory parameters
        (coercivity, remanent polarization, piezoelectric
        coefficients $d_{ij}$, \ldots), classified by property
        nature. Cannot be replaced by QUDT alone. The \leveltwo{}
        portion (\Cref{tab:l2_modules}) holds the tensor-component
        sub-properties. \\
    \texttt{process}
      & 263
      & ProcessStep hierarchy (Mixing, Calcination, Sintering,
        Poling, Patterning, \ldots); workflow primitives are
        domain-stable across material classes. \\
    \texttt{investigation}
      & 224
      & Investigation, Study, Assay, Sample, Batch, SampleState,
        Provenance --- the ISA lab-notebook skeleton, plus DCL
        0--5 and ExperimentStatus lifecycle (planned $\to$
        in-progress $\to$ completed $\to$ archived). \\
    \texttt{equipment}
      & 218
      & Apparatus (furnaces, ball mills, spray dryers, XRD
        diffractometers, \ldots). Cross-domain anchor for
        asset-management and procurement tools. \\
    \texttt{supplier}
      & 137
      & Supplier and Specification --- provenance of feedstock,
        commercial identifiers, lot numbers, certificate-of-analysis
        anchors. Material-agnostic procurement vocabulary. \\
    \texttt{representation}
      & 89
      & Data-shape vocabulary: \texttt{ScalarValue},
        \texttt{Vector}, \texttt{Matrix}, \texttt{Tensor} (with
        rank-specific subclasses), \texttt{TimeSeries},
        \texttt{ProcessCurve}, \texttt{ThermalProfile}.
        \leveltwo{} instantiates these shapes with concrete axes
        (e.g.\ piezoelectric hysteresis loops, phase diagrams). \\
    \texttt{instrument}
      & 82
      & Analytical instruments and their detection components
        (\texttt{NMR\_Spectrometer}, laser-flash apparatus, Raman
        spectrometer, diffraction detectors, AFM tips, \ldots).
        Split from \texttt{equipment} so that instrument-level
        modelling --- with manufacturer, signal chain, and
        detector granularity --- is separable from the broader
        durable-asset taxonomy. \\
    \texttt{identifier}
      & 63
      & External-identifier types (CAS, InChI, ICSD, Pearson,
        Wikidata, lot number, \ldots). A three-tier hierarchy
        (mandatory / recommended / optional, ${\sim}45$ types in
        total) is SHACL-validated per material class. \\
    \texttt{simulation}
      & 40
      & Simulation methods (DFT, MD, phase field, FEM, CALPHAD)
        and software anchors. Treated as a sibling of
        \texttt{process}: a simulation step is a step in the
        workflow, not an afterthought. \\
    \texttt{documents}
      & 29
      & Document taxonomy spanning lab/industry artefacts
        (\texttt{TechnicalManual}, \texttt{ProductDataSheet},
        \texttt{SafetyDataSheet}, \texttt{CertificateOfAnalysis},
        \texttt{CalibrationCertificate}) plus the
        publication subtypes bridged to FaBiO/CiTO. Required so
        that LiteratureExtraction can be modelled as a first-class
        \texttt{prov:Activity}. \\
    \texttt{consumables}
      & 22
      & Single-use lab consumables, dominated by crucibles and
        boats (alumina, zirconia, platinum, \ldots). Material-
        agnostic procurement vocabulary that does not fit the
        durable-asset framing of \texttt{equipment}. \\
    \texttt{diffraction}
      & 17
      & Diffraction-specific concepts (\texttt{DiffractionReflection},
        \texttt{OrientationReflection},
        \texttt{ResidualDensityAnalysis},
        \texttt{StructureRefinement},
        \texttt{ReflectionStatistics}). Kept separate from
        \texttt{measurement} because the diffraction sub-tree
        carries its own analytic methods and refinement metadata
        beyond the generic measurement-step vocabulary. \\
    \texttt{environment}
      & 17
      & Process and storage atmospheres (air, inert, argon,
        nitrogen, \ldots) annotated with composition. Cross-cutting
        and material-agnostic, so factored out of
        \texttt{process}. \\
    \texttt{material}$\dagger$
      & 9
      & Material superclass plus four substance-based subclasses
        (Metal, NonmetallicInorganic, Organic, Composite) as the
        agnostic taxonomy frame; concrete materials (\bntbt{},
        KNN, \ce{Al2O3}, ferritic ceramics, \ldots) live in
        \leveltwo{} under these branches. Process roles are
        modelled separately in the \texttt{oco-role} module
        (\Cref{tab:l2_modules}). \\
    \texttt{rationale}
      & 5
      & DesignRationale + RationaleConfidence: meta-modelling for
        \emph{why} a sample/process/specification was chosen ---
        captures decision provenance, not just data provenance. \\
    \midrule
    \textbf{Total \levelone{}} & \textbf{2\,198} & \\
    \bottomrule
  \end{tabularx}
\end{table}

\textbf{A core \levelone{} pattern: literature and laboratory data
as semantically equivalent sources.} \oco{} treats
published-literature extractions and own-laboratory experiments as
first-class equivalent ABox contributors. Both are modelled as
\texttt{prov:Activity} instances and share the same provenance,
quality, and uncertainty annotations --- distinguished only by
their activity class (\texttt{Investigation} vs.\
\texttt{LiteratureExtraction}) and their verification status, not
by a separate vocabulary. A SPARQL query asking ``what is the
$d_{33}$ of \bntbt{}?'' returns both validated lab measurements
and extracted literature values, ranked by data-confidence level,
without the schema forcing one source to be preferred a priori.
This unification is the operational consequence of the equality
principle (\Cref{sec:positioning}, principle~5) and is the
architectural prerequisite for using large-scale literature
corpora alongside in-house experimentation; it also enables the
SIPOC-fragment data model (\Cref{sec:sipoc}) to accommodate
incomplete records uniformly across both source types.

The level is explicitly designed to be ELN-/LIMS-importable.
Beyond the module structure, four cross-cutting features make
\levelone{} operational for laboratory-data systems:

\begin{itemize}
\item \textbf{Data confidence levels (DCL 0--5):} a mandatory
      annotation on every \texttt{MeasurementResult} or
      \texttt{ExtractedValue}, ranging from \emph{Unknown} (DCL~0) to
      \emph{Validated} (DCL~5). This separates raw entries, peer-
      reviewed values, and gold-standard data within the same
      knowledge graph.
\item \textbf{ExperimentStatus lifecycle:} planned $\rightarrow$
      in-progress $\rightarrow$ completed $\rightarrow$ archived,
      with a separate \texttt{StatusArchived} terminal state.
      Required for any LIMS that distinguishes intent from execution.
\item \textbf{GUM uncertainty fields:} CombinedStandardUncertainty
      with the five datatype properties prescribed by JCGM~100:2008,
      attached to existing \texttt{MeasurementUncertainty} classes
      rather than introduced as a parallel hierarchy.
\item \textbf{LiteratureExtraction as first-class
      \texttt{prov:Activity}:} extraction from publications and
      patents is a tracked activity with its own VerificationStatus,
      so SIPOC-fragment data (\Cref{sec:sipoc}) carries its own
      provenance trail rather than being silently merged with
      laboratory data.
\end{itemize}

\textbf{Substitutability.} The level's existence as a separate
artefact is the architectural lever that distinguishes OCO from a
monolithic ceramics ontology: a sister project (metallurgy, polymer
science, pharmaceutics) imports \levelone{} unchanged together with
its own \leveltwo{}, replacing only the material-class-specific
content. This is the structural mechanism by which the
``${\sim}70$\,\% workflow duplication across PMD-consortium
projects'' problem identified in
\citet{norouzi_2024_landscape} is dissolved.

\subsection{Level 2 --- Material-Class-Specific Level}
\label{sec:level2}

\leveltwo{} contains the material-class-specific knowledge --- the
part of OCO that a sister project (metallurgy, polymer science,
pharmaceutics) would replace while keeping
\levelzero{}+\levelone{} unchanged. For ceramics it has two
internal facets: the \emph{material modules} themselves
(\Cref{sec:l2_modules}), and the \emph{seven-tier
mechanistic-explanation skeleton} (\Cref{sec:7layer}) that
structures how the material content can be queried for
cause-chain explanations rather than only for property values.

\subsubsection{Material Modules}
\label{sec:l2_modules}

For ceramics, \leveltwo{} decomposes into nine architecturally
significant modules (\Cref{tab:l2_modules}) that together cover the
structure-property-defect chain that distinguishes
functional-ceramic modelling from generic materials description.
Four additional smaller modules
(\texttt{dataquality}~17, \texttt{dimension}~14,
\texttt{application}~12, \texttt{calc}~10) carry support
vocabulary for the same level and total 53 further classes. A
fifth module, \texttt{oco-role}, holds the abstract
\texttt{Role} class plus 14 named individuals (\texttt{solvent},
\texttt{binder}, \texttt{precursor}, \texttt{reactant},
\texttt{dopant}, \texttt{plasticizer}, \ldots) that are attached
to process steps rather than to material classes --- the pattern
mentioned alongside the \texttt{material} module
(\Cref{tab:l1_modules}). The \texttt{oco-localstructure} module
listed below additionally contributes 7 \levelone{}-frame classes
(CoordinationPolyhedron root + family abstracts Tetrahedron /
Octahedron / Cuboctahedron) alongside its 12 \leveltwo{}
polyhedra.

\begin{table}[htbp]
  \centering
  \caption{The principal \leveltwo{} modules for ceramics (all
  under the \texttt{oco-*} namespace). \emph{Classes} counts the
  \leveltwo{}-level classes per module in the merged distribution.
  The modules marked $\dagger$ also carry \levelone{} content
  (\Cref{tab:l1_modules}). A sister project for a different
  material class would replace these modules while leaving
  \levelzero{}+\levelone{} untouched.}
  \label{tab:l2_modules}
  \footnotesize
  \begin{tabularx}{\linewidth}{@{}>{\raggedright\arraybackslash}p{1.8cm}r>{\raggedright\arraybackslash}X@{}}
    \toprule
    Module & Classes & Why this module is in \leveltwo{} \\
    \midrule
    \texttt{property}$\dagger$
      & 1\,018
      & Material-specific tensor-component sub-properties (Voigt
        components $d_{ij}$, $s_{ijkl}$, $\pi_{ij}$, \ldots) and
        their multi-axis classification (\Cref{sec:multiaxis}).
        Generated from family-templates plus outlier-override; the
        \levelone{} portion (\Cref{tab:l1_modules}) holds the
        laboratory backbone. \\
    \texttt{crystal}
      & 317
      & Crystal structure: 230 space groups, 32 point groups,
        Bravais lattices, Pearson symbols, plus functional-ceramic
        tags (centrosymmetric, polar, piezoelectric, pyroelectric,
        enantiomorphic). Space groups are universal physics, but
        the tags that connect symmetry to functional behaviour are
        ceramics-flavoured. A polymer ontology would replace the
        tag set with chain-conformation descriptors. \\
    \texttt{material}$\dagger$
      & 259
      & Concrete material classes (\bntbt{}, KNN, BCZT,
        \ce{Al2O3}, \ce{ZrO2}, ferritic high-performance ceramics,
        \ldots) and their composition relationships. The skeleton
        sits in \levelone{}; the catalogue that a sister project
        replaces with its own metals/polymers/actives lives here. \\
    \texttt{tensor}
      & 191
      & A \texttt{MaterialTensor} root plus 190 tensor-class
        sub-roots (Voigt families, polarization, conductivity,
        elasticity, piezo-, pyro-, magnetoelectric, \ldots), each
        annotated with tensor order, symmetry class, and
        axial/polar flag. The structural anchor for the Neumann
        engine (\Cref{sec:neumann_phase}). \\
    \texttt{element}
      & 135
      & Periodic-table classes with element-specific properties
        (atomic number, electronegativity, common oxidation states,
        ionic radii). Bridged to Wikidata. Elements sit in
        \leveltwo{} because element subsetting is
        material-specific: ceramic-relevant elements (Ti, Zr, Pb,
        Bi, Na, K, \ldots) differ from those that drive metals
        (Fe, Al, Cu, Ni) or polymers (C, H, O, N). \\
    \texttt{defect}
      & 91
      & Defect chemistry. Kröger-Vink notation for ionic solids
        \citep{kroeger_vink_1956} (\(V_O^{\bullet\bullet}\),
        \(Mg_{Si}''\)) and Brouwer-diagram modelling of defect
        equilibria. A metallurgy ontology would replace this
        wholesale with Burgers-vector dislocation notation; a
        polymer ontology with chain-defect descriptors. \\
    \texttt{phase}
      & 73
      & Phases and phase transitions: solid solutions, morphotropic
        phase boundaries, ferroelectric/paraelectric transitions,
        plus concrete phase regions for \bntbt{} (four with
        $(T, x)$ bounds) and NiCuZn ferrite (Curie transition).
        The substrate that the phase-state coupling
        (\Cref{sec:neumann_phase}) consumes. \\
    \texttt{composite}
      & 13
      & Composite topology. Newnham connectivity
        \citep{newnham_1978_connectivity} (0-0, 0-3, 1-3, 2-2,
        3-3, \ldots) for two-phase functional-ceramic composites.
        Small but architecturally important: the connectivity
        notation is what makes piezoelectric-composite reasoning
        possible at all. \\
    \texttt{layer}
      & 6
      & The five material-abstraction layers (Atomic, Crystalline,
        Microstructural, Mesoscopic, Macroscopic) plus their
        Layer root. Each tier is a distinct tuple of
        \emph{state variables}, \emph{governing laws}, and
        \emph{characteristic length scale} --- not a spatial
        partition (\Cref{sec:multiaxis}). \\
    \texttt{localstructure}$\dagger$
      & 12
      & Concrete coordination polyhedra (TiO\textsubscript{6},
        BiO\textsubscript{12}, BaO\textsubscript{12},
        SiO\textsubscript{4}, ZnO\textsubscript{4},
        NbO\textsubscript{6}, FeO\textsubscript{6},
        NiO\textsubscript{6}, \ldots) inheriting from the
        \levelone{} family abstracts. Closes the mechanistic chain
        \emph{composition $\to$ local geometry $\to$ symmetry $\to$
        allowed tensor components}
        (\Cref{sec:neumann_phase}). \\
    \midrule
    \textbf{Total} (principal \leveltwo{}) & \textbf{2\,115} & \\
    \bottomrule
  \end{tabularx}
\end{table}

A cross-cutting feature spans \leveltwo{} and \levelone{}:
the \textbf{coupled-effect family} (32 effect classes:
piezoelectric, pyroelectric, electrostriction, magnetostriction,
magnetoelectric, thermoelectric, Verdet, Cotton-Mouton, plus 24
diagonal/cross effects following Nye's convention
\citep{nye_1985}). The effect classes themselves are physical
quantities and live in \texttt{oco-property} at \levelone{}; the
symmetry constraints that connect each effect to its allowed
crystallographic point groups live at \leveltwo{} via the
\texttt{oco-tensor} module and the reified
\texttt{oco:NeumannConstraint} instances generated programmatically
by the Neumann tensor engine (\Cref{sec:neumann_phase}). This
split keeps the universal physical-quantity hierarchy in the
agnostic level while the material-bound symmetry constraints stay
where they belong --- and replaces what would otherwise be hundreds
of manually curated cross-axioms with a single algorithm.

\textbf{Substitutability.} The nine principal modules in
\Cref{tab:l2_modules} are the surface that a sister project
replaces. The ferritic-high-performance-ceramics implementation
currently in development is the first concrete test of this
substitutability within the ceramics family; cross-domain
substitutability (metallurgy, polymers) is on the validation
roadmap (\Cref{sec:discussion}).

\subsubsection{Seven-Tier Mechanistic-Explanation Skeleton}
\label{sec:7layer}

The material-specific content of \leveltwo{} is organised
internally by a seven-tier mechanistic-explanation skeleton that
distinguishes an authoritative materials ontology from a vocabulary
that only describes. Every material property recorded in \oco{}
sits within a chain of physical mechanisms; for industrial use
that chain must be queryable from the same knowledge graph as the
property value itself. \oco{} formalises this chain as a fixed
skeleton of seven explanation tiers, each answering one
fundamental materials-science question and contributing dedicated
modules and (where applicable) an external reference-data cache
(\Cref{tab:7layers}).

\paragraph{Why ``tiers'' --- five material-abstraction layers vs.\ seven mechanistic-explanation tiers}
We deliberately call the seven mechanistic-explanation stages
\emph{tiers}, not layers, to keep them terminologically distinct
from the five \emph{material-abstraction layers} (Atomic,
Crystalline, Microstructural, Mesoscopic, Macroscopic) carried by
the \texttt{oco-layer} module listed in \Cref{tab:l2_modules}. The
two concepts answer different questions. The material-abstraction
layers describe the \emph{scale and governing-law regime} at which
a property is observed --- each is a distinct tuple of state
variables and governing laws used as a classification axis for
parameters (\Cref{sec:multiaxis}). The mechanistic-explanation
tiers describe the \emph{causal mechanism stack} that links
composition through symmetry, energy, thermodynamics, kinetics,
microstructure, defects, and bonding to the observed property.
\emph{Microstructure} appears in both lists by name only: in the
material-abstraction sense (layer) it is a description scale
(grain-sized features); in the mechanistic-explanation sense
(tier) it is the cause-chain stage that supplies polycrystalline
averaging, Hall--Petch coupling, and
texture effects.

\begin{table}[htbp]
  \centering
  \caption{The seven mechanistic-explanation tiers in \oco{},
  with implementing modules and external reference-data caches.
  Each tier answers one fundamental question about a crystalline
  ionic oxide. Caches are pinned to upstream versions in
  \texttt{bridge/external\_versions.yaml} per the cache-pattern
  principle (\Cref{sec:arch_principles}).}
  \label{tab:7layers}
  \scriptsize
  \begin{tabularx}{\linewidth}{@{}c>{\raggedright\arraybackslash}p{1.4cm}X>{\raggedright\arraybackslash}p{2.7cm}>{\raggedright\arraybackslash}p{2.2cm}@{}}
    \toprule
    \# & Tier & Question answered & Implementing modules & External cache \\
    \midrule
    1 & Symmetry         & Which crystal structure, which symmetry operations? Point-group theory (Neumann's principle) governs which tensor components are allowed; subgroup-relations drive phase transitions; Wyckoff positions give dopant and defect sites. & \texttt{oco-symmetry} (11) atop \texttt{oco-crystal}, \texttt{oco-tensor} & 1\,731 Wyckoff positions (pyxtal); 1\,934 bond-valence parameters (IUCr Brown 2020) \\
    2 & Energy / DFT     & Which electronic structure, which phonon modes, which Born effective charges? First-principles ground for intrinsic properties without empirical fitting; soft-mode analysis identifies phase-transition mechanisms. & \texttt{oco-energy-dft} (33 classes + 7 properties) & $\sim$155\,000 Materials Project DFT entries \\
    3 & Thermo / CALPHAD & Which phases are stable at $(T, p, x)$? Sublattice models and Redlich--Kister polynomials govern Gibbs-energy surfaces; sintering windows, miscibility gaps, and morphotropic phase boundary locations follow. & \texttt{oco-thermo} (25, incl.\ MPB + chemical-potential classes) atop \texttt{oco-phase} & --- \\
    4 & Kinetics         & How fast do diffusion, reaction, sintering, switching, and aging proceed? Fick, Arrhenius, JMAK, Hillert grain growth, Cahn--Hilliard spinodal decomposition, domain-wall mobility. & \texttt{oco-kinetics} (Fick, Arrhenius, JMAK, Hillert, Cahn--Hilliard, DW mobility, aging) & --- \\
    5 & Microstructure   & Which grain size, texture, and grain-boundary character? Polycrystalline averaging through Voigt--Reuss--Hill, Mori--Tanaka and Hashin--Shtrikman bounds; Hall--Petch coupling to mechanical response. & \texttt{oco-microstructure} (26: VRH, MT, HS, Hall--Petch, CSL grain boundaries) & --- \\
    6 & Defect chemistry & Which point defects, in which concentrations, with which charge compensation? Kröger--Vink notation, Brouwer diagrams, donor / acceptor / iso-valent / amphoteric doping strategies. & \texttt{oco-defect} extended (+48 Kröger--Vink, Brouwer, doping classes) & --- \\
    7 & Bonding chemistry & Which electron configuration, which hybridisation, which bonding character (ionic / covalent / metallic)? Pauling rules, HSAB classification, crystal-field theory, lone-pair stereo-activity. & \texttt{oco-bonding} (28: HSAB, lone-pair, hybridisation, Pauling rules, crystal field) & 497 Shannon ionic radii; 91 Pauling electronegativities \\
    \bottomrule
  \end{tabularx}
\end{table}

\paragraph{Generic skeleton, not material-specific}
All seven tiers apply to every \emph{crystalline ionic oxide} ---
the central material class for functional ceramics
(ferroelectrics, magnetics, ion conductors, high-$T_c$
superconductors). \bntbt{} is the pilot material on which the
skeleton was concretised; it is not the reason the skeleton has
these seven tiers. The same skeleton applies without
restructuring to NiCuZn ferrite, yttria-stabilised zirconia,
\ce{BaTiO3}, and any related material domain --- only the
per-tier instance content changes.

\paragraph{External reference-data caches}
Each tier that draws on a large external reference corpus does
so through the cache pattern formalised as Principle~5 in
\Cref{sec:arch_principles}: a versioned local snapshot rather
than embedded TBox classes. The per-tier cache totals appear in
the right-most column of \Cref{tab:7layers}; all caches are
SHA-pinned in \texttt{bridge/external\_versions.yaml}. The OWL
TBox stays compact; the externally curated data scales
independently.

\paragraph{Cross-tier validation}
The chain of explanation is auditable. Cross-tier references are
formalised as seven SHACL NodeShapes
(\path{shapes/m72_cross_layer_shapes.ttl}) that flag
incomplete annotations: a morphotropic phase boundary
(Tier 3 Thermo) must reference a domain-state set
(Tier 1 Symmetry); a soft mode (Tier 2 Energy/DFT) must
reference a subgroup relation (Tier 1 Symmetry); a Brouwer
diagram (Tier 6 Defect chemistry) must reference a
chemical-potential diagram (Tier 3 Thermo); an aging-kinetics
model (Tier 4 Kinetics) must reference a Kröger--Vink defect
(Tier 6 Defect chemistry); a Hall--Petch model
(Tier 5 Microstructure) must reference a grain-boundary
classification (Tier 6 Defect chemistry, via grain-boundary
defect types); a Born effective charge (Tier 2 Energy/DFT) must
reference a bonding-character annotation (Tier 7 Bonding
chemistry); a spinodal decomposition (Tier 3 Thermo) must
reference a Cahn--Hilliard model (Tier 4 Kinetics). The
constraints are deliberately lightweight (\texttt{sh:Warning}
severity, not \texttt{sh:Violation}) because cross-tier
annotation grows evolutionarily; missing references should
attract modeller attention, not block the pipeline.

\paragraph{End-to-end pilot}
The \bntbt{} pilot ABox
(\path{examples/abox_pilots/bnt_bt_pilot.ttl}) carries
concrete instances on each of the seven tiers with explicit
cross-tier annotations: a soft mode at the $\Gamma$ point linked
to the R3c$\to$P4mm subgroup relation; a Born effective charge
$Z^*_{33}(\text{Ti}) \approx +7.2$ explained by Bi-6s$^2$
lone-pair stereo-activity; a morphotropic phase boundary at
$x_\text{BT} \approx 0.06$ linked to the R3c/P4mm/Cm domain-state
set; an aging-kinetics model linked to the
V$_\text{O}^{\bullet\bullet}$ / Mn$_\text{Ti}^{''}$ defect-dipole
chemistry; a Hall--Petch parameterisation with $\sigma_0 \approx
3\,\text{GPa}$, $k_{HP} \approx 0.8\,\text{MPa}\sqrt{\text{m}}$
at $d = 2\,\mu\text{m}$; and a complete DFT-workflow provenance
(PBEsol functional, PAW pseudopotentials, $8\times8\times8$
Monkhorst--Pack mesh, $600\,\text{eV}$ cutoff). The pilot
demonstrates that the skeleton holds together operationally, not
only conceptually --- a query for ``why does \bntbt{} achieve
$d_{33} \approx 580\,\text{pC/N}$ at the MPB'' traverses all
seven tiers along the cross-tier annotation chain.

\subsection{Level 3 --- Categorical Reasoning Level}
\label{sec:level3}

\levelthree{} carries the categorical (boolean-logical) reasoning
that crosses module boundaries --- the truths that hold
\emph{always}, \emph{never}, or \emph{under condition X}, and that
have no natural home in any single \levelone{}/\leveltwo{} module
because they connect classes from several. Quantitative
mathematics (functions, scaling laws, calibration curves) is
\emph{not} part of \levelthree{}; it lies outside this release
entirely (\Cref{sec:scope_boundary}). \levelthree{} ships as two TTL distributions,
\texttt{oco-domain.ttl} (categorical axioms and compatibility
tables) and \texttt{oco-routes.ttl} (ordered process-step templates
and their state-sequence entries), compiled from YAML axiom sources
under the \texttt{axioms/} directory. Their combined contribution
to the merged distribution, measured as the ROBOT
\texttt{measure} \citep{jackson_2019_robot} delta
between \texttt{oco\_merged\_l12.ttl} (\levelone{}+\leveltwo{})
and \texttt{oco\_merged.ttl} (\levelone{}+\leveltwo{}+\levelthree{}),
is 1\,400 OWL axioms / 404 logical axioms / 248 ABox axioms. The
content decomposes into four families (\Cref{tab:l3_families}).

\begin{table}[htbp]
  \centering
  \caption{The four record families in \levelthree{}.
  \emph{Records} counts source-level entries; each record typically
  emits several OWL axioms or reified-constraint triples in the
  compiled distribution.}
  \label{tab:l3_families}
  \footnotesize
  \begin{tabularx}{\linewidth}{@{}>{\raggedright\arraybackslash}p{2.4cm}r>{\raggedright\arraybackslash}X@{}}
    \toprule
    Family & Records & Why this family is in \levelthree{} \\
    \midrule
    Domain axioms
      & 74
      & Categorical constraints that link classes from two or more
        modules. Symmetry-effect axioms (Pyroelectricity requires a
        polar point group), lifecycle constraints (PoledBody may
        transition only to AnnealedBody, AgedBody, or
        FatiguedBody), property-disjointness facts
        (\texttt{agrees\_with}~$\bowtie$~\texttt{disagrees\_with}),
        and partition-disjointness axioms over the 10
        material-abstraction layers (\Cref{tab:l2_modules}) and the
        21 crystal-system root classes. These do not belong in any
        one module because they cross
        Property$\times$Crystal,
        Process$\times$State, or
        Property$\times$Property boundaries. \\
    Route templates
      & 14
      & Ordered process-step sequences per (synthesis route
        $\times$ application domain), e.g.\
        SolidState\,$\times$\,Piezoelectric prescribes Mixing
        $\to$ DryMilling $\to$ Calcination $\to$ \ldots $\to$
        Poling as the canonical chain. Required to validate
        whether a recorded ABox process chain is a legal instance
        of a known route, and to answer comparability questions
        across publications. Each template unfolds into ordered
        step and state-sequence records (129 steps and 43
        state-sequence entries in total). \\
    Compatibility entries
      & 56
      & Status classification of the route\,$\times$\,domain
        matrix (\emph{standard}, \emph{rare}, \emph{incompatible}),
        e.g.\ ThinFilmRoute\,$\times$\,Structural is incompatible.
        Lets the ontology distinguish ``unusual but valid'' from
        ``rule violation''. \\
    L3-justifying CQs
      & 144 of 163
      & Each \levelthree{} axiom is justified by at least one
        published competency question. Of the full CQ catalogue
        (count in \Cref{sec:implementation}), 144 entries map to
        \levelthree{} axioms; the remainder cover
        \levelone{}/\leveltwo{} patterns. 52 CQs carry an
        executable SPARQL test against gold-standard ABoxes
        (52/52 PASS). Questions are tagged with one of ten
        reasoning \emph{areas} (tensor symmetry, phase state,
        route, lifecycle, multi-axis, interoperability, quality,
        crystal structure, BNT-BT, and the NCZF placeholder),
        making the reasoning scope auditable per area. \\
    \midrule
    \textbf{Total} & \textbf{288} & source-level entries; their
    compiled contribution to \texttt{oco\_master\_full.ttl} is
    1\,400 OWL axioms / 404 logical (ROBOT-measured delta over
    \levelone{}+\leveltwo{}) \\
    \bottomrule
  \end{tabularx}
\end{table}

A characteristic example is the symmetry-effect axiom
\begin{equation}
\textit{Pyroelectricity} \sqsubseteq \textit{requires\_pointgroup} \mathbin{\textit{only}} \textit{PolarPointGroup},
\end{equation}
which encodes the categorical fact that a material can show
pyroelectricity only if its crystallographic point group is one of
the ten polar groups. This is the kind of axiom whose evaluation
requires the universal restriction (\texttt{only}) construct ---
which is exactly the construct that lifts the OCO-with-\levelthree{}
profile from OWL\,2\,EL to OWL\,2\,DL
(\Cref{tab:expressivity}). Consumers without this reasoning need
neither load \levelthree{} nor pay the OWL\,2\,DL reasoner cost.

The decisive architectural feature of \levelthree{} is not the
axioms themselves but their \emph{separation} from the rest of the
schema and their binding to published competency questions. This
makes the scope of the ontology --- which questions it intends to
answer --- explicit, auditable, and testable: a CQ that loses its
SPARQL test is a visible regression, and a proposed new axiom that
cannot point to a CQ has no architectural justification for being
admitted.

\subsection{The Material / Compliance Audience Axis}
\label{sec:audience_axis}

A second classification axis is independent of the level
hierarchy: every module carries an \texttt{audience} marker, with
value \texttt{material}, \texttt{compliance}, or both. Twenty-nine
modules constitute the material-audience-only set --- the
materials-science vocabulary proper, including
\texttt{oco-localstructure} (the mechanistic bridge from
composition through coordination polyhedra to symmetry,
\Cref{sec:7layer}). Eleven modules form the compliance-audience-only
set and carry the EU-regulatory and value-chain vocabulary
detailed in \Cref{sec:vertical_extension}: Life Cycle Assessment,
CSRD/ESRS reporting, supply-chain due diligence (CSDDD),
packaging (PPWR), carbon-border adjustment (CBAM), right-to-repair
(R2R), the AI Act, Manufacturing-X identifier and traceability
infrastructure, regulated substances, recycling, and
Safe-and-Sustainable-by-Design. Four further modules are
\emph{dual-audience} --- they serve both sides simultaneously:
\texttt{oco-format} (industrial data formats), \texttt{oco-odrl}
(W3C ODRL policy and Verifiable Credentials trust layer),
\texttt{oco-time-event} (temporal extents per W3C Time), and
\texttt{oco-automation} (SiLA~2.0 laboratory automation). These
appear in \Cref{tab:compliance_modules} marked with $\dagger$.

The two audiences are independent of the four levels. Every
compliance module has its own
\levelone{}/\leveltwo{}/\levelthree{} internal structure, just as
the material modules do. A consumer who needs only the material
core can ignore the compliance audience entirely; the
\texttt{audience} marker drives module-selection profiles in the
distribution. A consumer building a Manufacturing-X data pipeline
loads both audiences. The independence is what makes the same
architectural primitive --- modular layering --- absorb a second
class of requirements without restructuring the material core.


\section{Reference Implementation: OCO}
\label{sec:implementation}

\subsection{Quantitative Description}

The OntoCrafter Ceramics Ontology (\oco{}) is the reference
implementation of the architecture introduced in
\Cref{sec:architecture} --- two independent classification axes
(level and audience) plus the seven-tier mechanistic-explanation
structure of the material level. Metrics for the v0.94 release ---
the version entering productive practice --- are measured with
ROBOT~1.9.10 \citep{jackson_2019_robot} and \texttt{rdflib} on
the merged distribution \texttt{distribution/oco\_merged.ttl}
extended by the Neumann-constraint sub-distribution:

\begin{itemize}
\item \textbf{Classes:} 5\,196 named, distributed across 44
      class-bearing modules (45 in the namespace registry). 29
      modules carry the material audience and 15 carry the
      compliance audience, four of them dual-audience
      (\Cref{sec:vertical_extension}). Every module carries its
      own \levelone{}/\leveltwo{}/\levelthree{} internal
      structure independent of the audience axis, and every
      material-audience module sits on one or more of the seven
      mechanistic explanation tiers (\Cref{sec:7layer}).
\item \textbf{Properties:} 1\,674 total ---
      574 ObjectProperties, 1\,051 DatatypeProperties, and 49
      AnnotationProperties.
\item \textbf{Axioms:} 167\,348 OWL/RDFS axioms in the merged
      distribution, of which 40\,454 are logical axioms (the
      remainder are annotations and bare entity declarations);
      per ROBOT \texttt{measure} \citep{jackson_2019_robot}. The
      largest single contribution is the 5\,920 reified
      \texttt{oco:NeumannConstraint} instances --- one per
      (tensor~$\times$~point-group) pair,
      \Cref{sec:neumann_phase}.
\item \textbf{Bridges:} 11 \levelzero{} targets with substantial
      class-level coverage ($\geq$14 anchors each,
      \Cref{tab:l0_bridges}); a further 829 explicit
      cross-ontology mappings across 40 sections in
      \texttt{bridge\_mappings.yaml} and \texttt{mwo\_mappings.yaml}
      cover the material-side standards (PMDco, PROV-O, NFDIcore,
      MADO, OBI, CIF Core, EMMO sub-modules, QUDT, Croissant,
      KnowNow, SmaDi), the compliance-side standards
      (EN 15804, EU PEF, ecoinvent, BONSAI, EFRAG XBRL, CRMA,
      NZIA, PACT, Catena-X, CSDDD-UNGP, CBAM, R2R, AI Act,
      AAS, \ldots), and the policy/trust layer (W3C ODRL,
      W3C Verifiable Credentials, SiLA) ---
      \Cref{sec:interop_extended}, \Cref{sec:vertical_extension}.
\item \textbf{Competency questions:} 163 published, each tagged
      with one of ten reasoning \emph{areas} (tensor symmetry,
      phase state, route, lifecycle, \ldots); 52 carry
      executable SPARQL test queries against gold-standard ABoxes
      (52/52 PASS).
\item \textbf{SHACL shapes:} 1\,172 mandatory shapes validated via
      \texttt{pyShacl}; dominated by the machine-generated Neumann
      (768) and CIF-Core (270) subsets, complemented by 7
      Cross-Tier NodeShapes (\Cref{sec:7layer}) and per-domain
      hand-curated sets for material-class identifiers, MPB
      coexistence, and the compliance audience.
\item \textbf{Architecture decision records:} 132, including
      explicit negative decisions for rejected bridge targets and
      rejected modelling alternatives.
\item \textbf{External reference-data caches:} 5 productive
      caches under the seven-tier explanation skeleton
      (\Cref{sec:7layer}), pinned to upstream versions in
      \texttt{bridge/external\_versions.yaml}: Shannon ionic
      radii (497), IUCr bond-valence parameters (1\,934), pyxtal
      Wyckoff positions (1\,731), Materials Project DFT corpus
      ($\sim$155\,000), Pauling electronegativities (91).
\item \textbf{Expressivity:} the full distribution sits in
      OWL\,2\,DL; the \levelzero{}+\levelone{}+\leveltwo{} bundle
      (without \levelthree{}) reduces to OWL\,2\,EL ---
      \Cref{tab:expressivity} gives the underlying ROBOT
      measurements.
\end{itemize}

\paragraph{Comparison with PMDco and EMMO}
To place these metrics in landscape context,
\Cref{tab:metric_comparison} compares \oco{}~v0.94 against
PMDco~v3 (\oco{}'s primary mid-level bridge target,
\Cref{sec:level0}) and EMMO Crystallography (the EMMO sub-module
closest in scope to ceramics). The base-metric values for the
comparators are taken from Norouzi et al.\
\citep{norouzi_2024_landscape}, Tab.~9. \oco{} is roughly an
order of magnitude broader than PMDco in classes and properties
and almost two orders of magnitude broader in axioms; this is
expected because PMDco is a mid-level vocabulary that \oco{}
extends downward into a full material domain. EMMO is federated
across 40+ sub-modules, each with its own per-module metrics; we
report the Crystallography sub-module as the most analogous
single point of comparison.

\begin{table}[htbp]
  \centering
  \caption{Metric comparison \oco{}~v0.94 against PMDco~v3 and
  EMMO Crystallography, two reference points from the MSE
  ontology landscape. Class, property, axiom, and
  annotation-axiom counts for PMDco and EMMO Crystallography are
  taken from \citet{norouzi_2024_landscape}, Tab.~9; ``---''
  marks metrics not reported as release-time statistics in that
  survey.}
  \label{tab:metric_comparison}
  \footnotesize
  \begin{tabularx}{\linewidth}{@{}>{\raggedright\arraybackslash}X>{\raggedleft\arraybackslash}p{1.8cm}>{\raggedleft\arraybackslash}p{1.6cm}>{\raggedleft\arraybackslash}p{2.0cm}@{}}
    \toprule
    Metric & \oco{}~v0.94 & PMDco~v3 & EMMO Cryst. \\
    \midrule
    Classes                              & 5\,196    &   264  &    61 \\
    Object Properties                    &    574    &    36  &     5 \\
    Data Properties                      & 1\,051    &     9  &     1 \\
    Total Axioms                         & 167\,348  & 2\,154 &   357 \\
    \quad of which Annotation Axioms     & 126\,894  & 1\,454 &   175 \\
    \midrule
    Reified Neumann constraints          & 5\,920    &    --- &   --- \\
    Cross-ontology bridges               &    829    &    --- &   --- \\
    SHACL shapes                         & 1\,172    &    --- &   --- \\
    Competency questions / SPARQL tests  & 163 / 52  &    --- &   --- \\
    \bottomrule
  \end{tabularx}
\end{table}

\subsection{Distribution Variants}

The OCO distribution is structured so that consumers select only the
depth they need (see \Cref{tab:distribution}). Both bundle files
---~\path{oco_master.ttl} and \path{oco_master_full.ttl}~--- are thin
\texttt{owl:Ontology} hulls carrying only \texttt{owl:imports}
directives; the actual content lives in 17~per-module
\path{oco-<modul>.ttl} files plus \path{bridge.ttl}. The two master
variants are \emph{alternatives}, not additive: importing the full
variant subsumes the standard one.

\begin{table}[htbp]
  \centering
  \caption{\oco{} distribution choice as a function of consumer
  profile. Reasoner column gives the minimal sufficient profile for the
  axioms actually present (see \Cref{tab:expressivity}).}
  \label{tab:distribution}
  \small
  \begin{tabularx}{\linewidth}{@{}>{\raggedright\arraybackslash}p{3.3cm}>{\raggedright\arraybackslash}Xl@{}}
    \toprule
    Consumer profile & Files to load & Reasoner \\
    \midrule
    PMDco / QUDT integration only
      & \texttt{bridge.ttl}
      & RDFS \\
    LIMS/ELN, no material focus
      & selected \levelone{} modules + \texttt{bridge.ttl}
      & RDFS \\
    Ceramics materials scientist
      & \texttt{oco\_master.ttl} (\levelzero{}+\levelone{}+\leveltwo{})
      & OWL\,2\,EL \\
    Routes / symmetry / lifecycle reasoning
      & \texttt{oco\_master\_full.ttl} (adds \levelthree{})
      & OWL\,2\,DL \\
    \bottomrule
  \end{tabularx}
\end{table}

The reasoner-profile assignment is empirical, not assumed: ROBOT
classifies the description-logic constructs actually used in each
bundle (\Cref{tab:expressivity}). The decisive contrast is that
adding \levelthree{} introduces universal restrictions ($\forall$,
\texttt{owl:allValuesFrom}, used in route-template completeness
constraints) and complex concept negation, both of which lift the
profile from OWL\,2\,EL into full OWL\,2\,DL. The
\levelzero{}--\levelone{}--\leveltwo{} bundle stays in
EL\nobreakdash-tractable territory --- a polynomial-time reasoner
(e.g.\ ELK) is sufficient. This is the multi-level-architecture promise
made operational: \emph{reasoning cost is a consequence of the level
chosen, not a property of the ontology}.

\begin{table}[htbp]
  \centering
  \caption{Description-logic constructs (ROBOT \texttt{measure}
  output) per bundle. The bold tokens in
  \texttt{oco\_master\_full.ttl} are the \levelthree{} additions
  that lift the profile out of OWL\,2\,EL.}
  \label{tab:expressivity}
  \footnotesize
  \begin{tabularx}{\linewidth}{@{}lX@{}}
    \toprule
    Bundle & Expressivity (constructs in use) \\
    \midrule
    \texttt{oco\_master.ttl}
      & \texttt{RRESTRERIF(D)} --- role hierarchy + restrictions,
        $\exists$, inverse, functional, datatypes \\
    \texttt{oco\_master\_full.ttl}
      & \texttt{RRESTR\textbf{C}\textbf{UNIV}RESTRERIF(D)} --- all of
        the above, plus \textbf{$\forall$ (universal restriction)} and
        \textbf{complex concept negation} \\
    \bottomrule
  \end{tabularx}
\end{table}

\subsection{Bilingual Definitions as Mandatory Field}

Each class carries both a German (\texttt{definition\_de}) and an
English (\texttt{definition\_en}) definition. This is enforced as a
hard schema-validator constraint. Among the surveyed MSE ontologies
\citep{norouzi_2024_landscape, zhang_2024_mgedkg, jalali_2023_msle},
none mandates bilingual definitions; \oco{} is to our knowledge the
only MSE ontology that does so as a hard constraint. This matters
specifically for German-speaking industrial partners, BMBF projects,
and the NFDI-MatWerk consortium.

\subsection{Source Lifecycle and Quality Assurance}

The distribution is generated from human-readable YAML sources rather
than maintained directly as Turtle, which keeps source files
git-diffable and review-friendly. Consistency of the distribution is
verified against standard tooling: HermiT for OWL-DL reasoning,
SHACL validators for shape constraints, and the OOPS! pitfall
scanner \citep{oops_pitfall_scanner} for ontology-engineering
hygiene. All build steps and quality checks are reproducible from
the source repository, ensuring that the published distribution can
be regenerated and audited independently.

\subsection{Authoritative Reasoning: Neumann Engine and Phase-State Coupling}
\label{sec:neumann_phase}

The authoritativeness principle (\Cref{sec:positioning},
principle~8) operationalizes the chain \emph{phase
diagram}~$\to$~\emph{active crystallographic phase}~$\to$~\emph{tensor-component
symmetry} as a two-step inference. Two implementation components
carry it.

\paragraph{Neumann tensor engine}
A dedicated module \texttt{oco-tensor} carries 190 material-tensor
root classes (Voigt families, polarization, conductivity,
elasticity, piezo-, pyro-, magnetoelectric, \ldots{}), each annotated
with tensor order, symmetry class, and the axial/polar flag. A
subprocess engine implements Neumann's principle as a
stabilizer-subgroup computation in \texttt{sympy} over the
32~crystallographic point groups, generating for every (tensor
$\times$ PG) pair the set of independent non-zero components. The
engine is validated against \citet[Table~9]{nye_1985} and IEEE
Standard~176 on a 30-case gold standard (30/30~pass) and emits
5\,920 reified \texttt{oco:NeumannConstraint} instances together
with 768 \texttt{sh:NodeShape} validators. A hash-keyed cache keeps
regeneration cheap: a full re-run after an engine fix completes in
5:30~min wall-time (cache-warm) versus 46~min for the cold first
run.

\paragraph{Phase-state coupling}
A symmetry library is only useful when the active point group can be
\emph{derived}, not assumed. Phase regions are therefore modeled as
sub-classes of \texttt{oco-phase:CrystalPhase} with temperature and
composition bounds expressed in the QUDT \texttt{QuantityValue}
pattern, and each region carries a \texttt{has\_point\_group}
restriction. Two reference phase diagrams are populated: \bntbt{}
(four regions, including the morphotropic phase boundary as a
3m+4mm coexistence region) and NiCuZn ferrite (ferrimagnetic /
paramagnetic across the Curie transition). A SPARQL-\texttt{INSERT}
pass over the sample ABox infers the active region from $(T, x)$,
propagates the point group, and triggers the Neumann constraints
--- without OWL reasoning, which cannot express numeric intervals.
End-to-end inference is gold-standard tested on six cases
(6/6~pass). Together, the two components answer the central
authoritativeness competency question --- ``which piezoelectric
coefficients are non-zero for \bntbt{} at room temperature near the
morphotropic phase boundary?'' --- without any manual point-group
assignment by the user.

\subsection{Multi-Axis Parameter Classification}
\label{sec:multiaxis}

Implements the facetedness principle (\Cref{sec:positioning},
principle~9). Every parameter class in \oco{} now carries three
independent classification axes simultaneously, each encoded as
\texttt{rdfs:subClassOf} so that an OWL\,2-DL reasoner navigates
them without bespoke join logic:

\begin{itemize}
\item \textbf{Role} (what the parameter does physically):
      StateVariable, ResponseParameter, TransportParameter,
      StructureParameter, StatisticalParameter,
      TopologicalParameter, FitParameter.
\item \textbf{Reference} (the semantic anchor):
      FundamentalConstant, MaterialParameter, InterfaceParameter,
      SpecimenParameter, ProcessParameter, MeasurementParameter,
      EnvironmentParameter.
\item \textbf{Layer} (abstraction level at which the parameter is
      defined): AtomicLayer, CrystallineLayer,
      MicrostructuralLayer, MesoscopicLayer, MacroscopicLayer ---
      five classes housed in a separate module
      \texttt{oco-layer}. Each carries its own \emph{state
      variables} (e.g.\ lattice parameters at the crystalline
      layer; effective permittivity at the mesoscopic layer) and
      \emph{governing laws} (Schr{\"o}dinger / DFT at the atomic
      layer; Maxwell-Garnett / Bruggeman at the mesoscopic layer).
      Layers are \emph{not} spatial partitions but coexisting
      descriptive views of the same material.
\end{itemize}

A bulk classification pipeline populates all three axes for the
$\sim$950 \texttt{MaterialParameter} subclasses through
family-default plus outlier-override rules. The tensor-rank and
symmetry-class annotations from \texttt{oco-tensor}
(\Cref{sec:neumann_phase}) provide a fourth independent hook that
connects the parameter taxonomy to the Neumann engine. The
resulting query granularity is illustrated by competency questions
of the form ``which response parameters at the crystalline layer
are non-zero for the trigonal point group~3m?'', which become
single \texttt{rdfs:subClassOf} traversals.

\subsection{Extended Interoperability and Predecessor Bridges}
\label{sec:interop_extended}

Realizes the adaptability, interoperability, and compatibility
principles (\Cref{sec:positioning}, principles~2, 3, 6) through a
disciplined bridge-expansion process. The discipline is captured in
a \emph{reuse-before-invention} checklist applied before any new
class or property is proposed: (a)~does \oco{} already have a
fitting class? (b)~external identifier? $\to$ subclass of
\texttt{oco-identifier:Identifier}; (c)~provenance? $\to$ PROV-O;
(d)~unit? $\to$ QUDT \texttt{QuantityValue}. In the most recent
interoperability wave, applying this checklist reduced eighteen
candidate new properties to zero and fifteen candidate classes to
five.

The expanded bridge inventory now contains 829~mappings across 40
sections in the \texttt{bridge\_mappings.yaml} and
\texttt{mwo\_mappings.yaml} single sources of truth:

\begin{itemize}
\item \textbf{Materials databases:} OPTIMADE, via a lookup table
      for the structure-search-API field vocabulary.
\item \textbf{Machine-learning datasets:} Croissant (MLCommons) for
      ML-ready dataset metadata.
\item \textbf{Electronic lab notebooks:} the ELN-Filetype standard,
      a cross-vendor RO-Crate-based export format adopted by
      14~ELN/LIMS systems --- bridging to the format covers the
      entire set without per-vendor mappings.
\item \textbf{EMMO sub-modules:} ISQ (the ISO/IEC~80000 quantity
      vocabulary, 93~mappings), chemistry, and materials, each in
      its own bridge section so that sub-module versioning can be
      tracked independently.
\item \textbf{Predecessor PMD ontologies:} KnowNow
      \citep{benhassine_2024_knownow} contributes 14~process- and
      property-class mappings (LTCC-multilayer-specific classes are
      deliberately excluded, as they belong to a material-specific
      route rather than the agnostic \levelone{}); SmaDi
      \citep{maas_2024_smadi} contributes 15~mappings restricted to
      the piezoelectric-ceramic subset, with the shape-memory and
      elastomer concepts excluded as out of ceramic scope.
\end{itemize}

Three thin application profiles ---
\path{oco_eln_profile.ttl},
\path{oco_materials_db_profile.ttl}, and
\path{oco_ml_profile.ttl} --- bundle the modules each consumer
class typically needs as \texttt{owl:imports} wrappers, avoiding
the maintenance load that subset-extracted distributions would
incur. External sources are versioned and cached locally, so that
a release change in any external target breaks only the
corresponding bridge file --- the adaptability principle made
operational.

\subsection{Compliance and Value-Chain Modules}
\label{sec:vertical_extension}

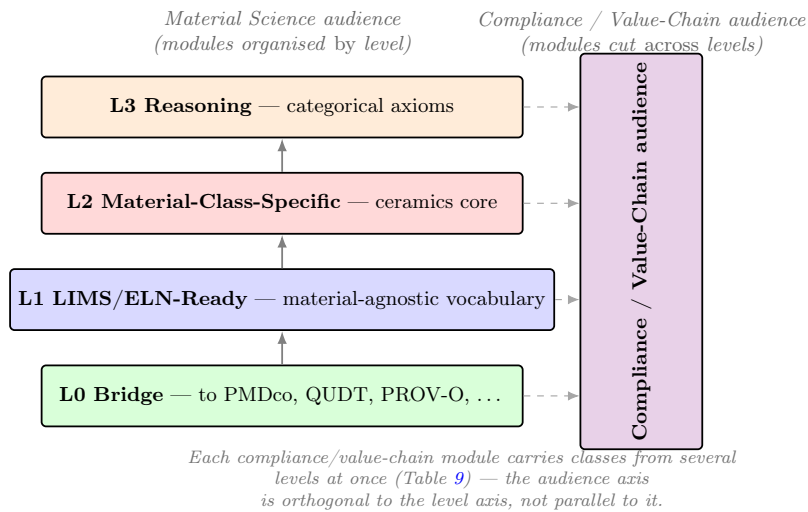
\begin{figure}[htbp]
  \centering
  \resizebox{0.85\linewidth}{!}{

\begin{tikzpicture}[
    layer/.style={
        draw, thick, rounded corners=2pt,
        minimum width=7.5cm, minimum height=0.95cm,
        font=\footnotesize, align=center
    },
    audience/.style={
        draw, thick, rounded corners=2pt,
        minimum width=1.9cm, minimum height=5.05cm,
        font=\footnotesize, align=center,
        fill=violet!18
    },
    grouplabel/.style={
        font=\footnotesize\itshape\color{black!60}, align=center
    },
    cut/.style={-{Latex[length=1.8mm,width=1.5mm]}, gray!70,
                line width=0.5pt, dashed}
]

\node[layer, fill=orange!15] (l3) at (0, 3.0) {%
    \textbf{L3 Reasoning} --- categorical axioms};
\node[layer, fill=red!15]    (l2) at (0, 1.5) {%
    \textbf{L2 Material-Class-Specific} --- ceramics core};
\node[layer, fill=blue!15]   (l1) at (0, 0.0) {%
    \textbf{L1 LIMS/ELN-Ready} --- material-agnostic vocabulary};
\node[layer, fill=green!15]  (l0) at (0, -1.5) {%
    \textbf{L0 Bridge} --- to PMDco, QUDT, PROV-O, \ldots};

\draw[-{Latex[length=2mm]}, thick, gray] (l0.north) -- (l1.south);
\draw[-{Latex[length=2mm]}, thick, gray] (l1.north) -- (l2.south);
\draw[-{Latex[length=2mm]}, thick, gray] (l2.north) -- (l3.south);

\node[audience] (comp) at (5.6, 0.75) {%
    \rotatebox{90}{\textbf{Compliance / Value-Chain audience}}};

\draw[cut] (l3.east) -- ([yshift=2.25cm]comp.west);
\draw[cut] (l2.east) -- ([yshift=0.75cm]comp.west);
\draw[cut] (l1.east) -- ([yshift=-0.75cm]comp.west);
\draw[cut] (l0.east) -- ([yshift=-2.25cm]comp.west);

\node[grouplabel] at (0, 4.15) {%
    Material Science audience\\
    (modules organised \emph{by} level)};
\node[grouplabel] at (5.6, 4.15) {%
    Compliance / Value-Chain audience\\
    (modules cut \emph{across} levels)};

\node[font=\scriptsize\itshape\color{black!55}, align=center]
    at (2.8, -2.8) {%
    Each compliance/value-chain module carries classes from several\\
    levels at once (\Cref{tab:compliance_modules}) --- the audience axis\\
    is orthogonal to the level axis, not parallel to it.};

\end{tikzpicture}}
  \caption{The compliance/value-chain audience as a cross-cutting
  strip independent of the \levelzero{}--\levelthree{} level
  hierarchy of the material-science core. The level hierarchy on
  the left (cf.\ \Cref{fig:level_stack}) is the organising
  principle for the material modules; the compliance/value-chain
  audience on the right is a single independent column that touches
  every level. Per-module level distributions are in
  \Cref{tab:compliance_modules}.}
  \label{fig:audience_axis}
\end{figure}

The audience axis introduced architecturally in
\Cref{sec:audience_axis} carries the value-chain and
EU-regulatory vocabulary on the compliance side. The same
construction-kit absorbs the vertical-convergence pressure of
\Cref{sec:introduction} --- EU-driven integration of materials,
manufacturing, supply-chain, and sustainability data --- without
restructuring the material core or the agnostic laboratory
level. \Cref{fig:audience_axis} captures the architectural
relationship: while the material-science audience is organised
\emph{by} level, the compliance/value-chain audience cuts
\emph{across} levels --- each compliance module carries classes on
several of \levelone{}, \leveltwo{}, and \levelthree{}
simultaneously (\Cref{tab:compliance_modules}).

\paragraph{Modules}
\Cref{tab:compliance_modules} lists the fifteen modules carrying
the compliance audience: eleven compliance-pure modules (LCA,
CSRD, CSDDD, PPWR, CBAM, R2R, AI Act, Manufacturing-X,
regulated substances, recycling, SSbD) and four dual-audience
modules (\texttt{format}, \texttt{odrl}, \texttt{time-event},
\texttt{automation}) shared with the material side. Each module
has its own \levelone{}/\leveltwo{}/\levelthree{} internal
structure --- the two axes (audience and level) are
independent.

\paragraph{Modules}
The compliance-audience modules cover the EU-regulatory wave that
will reshape materials-data practice over the next five years.

\begin{table}[htbp]
  \centering
  \caption{Modules carrying the compliance audience in \oco{}.
  Eleven are compliance-pure; four (marked $\dagger$) carry both
  the material and the compliance audience and are therefore
  shared between this table and the material modules. Each
  module has its own \levelone{}/\leveltwo{}/\levelthree{}
  structure (column \emph{L1/L2/L3}). The DPP column flags
  modules that contribute classes to the Digital Product Passport
  joining pattern.}
  \label{tab:compliance_modules}
  \footnotesize
  \begin{tabularx}{\linewidth}{@{}>{\raggedright\arraybackslash}p{1.0cm}>{\raggedleft\arraybackslash}p{0.5cm}>{\raggedleft\arraybackslash}p{1.6cm}>{\centering\arraybackslash}p{0.4cm}>{\raggedright\arraybackslash}X@{}}
    \toprule
    Module & Cls. & L1/L2/L3 & DPP & EU-regulatory anchor / role \\
    \midrule
    \texttt{lca}
      & 83 & 8 / 59 / 16 & $\bullet$
      & Life Cycle Assessment per ISO~14040/14044/14067,
        EN~15804+A2, EU PEF 2021/2279; bridges to ecoinvent,
        BONSAI. Material-DPP profiles attach here. \\
    \texttt{csrd}
      & 39 & 25 / 14 / -- &
      & Corporate Sustainability Reporting Directive (ESRS
        standards, double-materiality, reporting containers);
        EFRAG XBRL bridge. \\
    \texttt{mfgx}
      & 37 & 9 / 25 / 3 & $\bullet$
      & Manufacturing-X integration --- AAS asset administration
        shells, Catena-X BPN identifiers, IPCC AR climate factors,
        sector-DPPs (Battery / Construction CPR / Generic ESPR). \\
    \texttt{csddd}
      & 25 & 13 / 12 / -- &
      & Corporate Sustainability Due Diligence Directive (six-stage
        diligence process, supplier tiers); UNGP + ILO bridges. \\
    \texttt{ppwr}
      & 15 & 7 / 8 / -- &
      & Packaging and Packaging Waste Regulation (packaging
        function disjointness, sub-classes per use case). \\
    \texttt{regulation}
      & 14 & 3 / 11 / -- &
      & Cross-cutting regulated-substance vocabulary
        (recycling-inhibitor list, REACH-style anchors). \\
    \texttt{cbam}
      & 10 & 4 / 6 / -- &
      & Carbon Border Adjustment Mechanism (covered material
        categories: iron/steel, cement, electricity, \ldots). \\
    \texttt{r2r}
      & 8 & 4 / 4 / -- &
      & Right-to-Repair Directive (covered product categories:
        smartphone, tablet, large/small appliance). \\
    \texttt{aiact}
      & 8 & 4 / 4 / -- &
      & EU AI Act (risk tiers, conformity-assessment anchors). \\
    \texttt{recycling}
      & 6 & 4 / 2 / -- &
      & Recycling routes and recyclate streams. \\
    \texttt{ssbd}
      & 1 & 1 / -- / -- &
      & Safe-and-Sustainable-by-Design framework
        (Commission Recommendation 2022/2510), skeleton. \\
    \midrule
    \textbf{Subtotal} (compliance-pure) & \textbf{246} & 86 / 141 / 19 & & \\
    \midrule
    \texttt{format}$\dagger$
      & 42 & 5 / 37 / -- &
      & Industrial data-format vocabulary spanning ELN,
        simulation, LCA, Manufacturing-X, and document
        formats --- referenced by many of the
        compliance modules and by the material pipelines. \\
    \texttt{odrl}$\dagger$
      & 21 & 5 / 16 / -- &
      & W3C ODRL 2.2 usage-policy vocabulary and W3C Verifiable
        Credentials --- the trust and policy layer that
        compliance reporting and value-chain exchange rest on. \\
    \texttt{time-event}$\dagger$
      & 15 & 15 / -- / -- &
      & W3C Time-Ontology-aligned temporal extents, instants,
        and events --- required by reporting periods (CSRD/ESRS),
        validity windows, and audit-trail provenance. \\
    \texttt{automation}$\dagger$
      & 6 & 1 / 5 / -- &
      & SiLA 2.0 laboratory-automation standard --- shared
        between material instrumentation provenance and
        compliance traceability. \\
    \midrule
    \textbf{Subtotal} (dual-audience) & \textbf{84} & 26 / 58 / -- & & \\
    \midrule
    \textbf{Total} (compliance audience) & \textbf{330} & 112 / 199 / 19 & & \\
    \bottomrule
  \end{tabularx}
\end{table}

\paragraph{The joining concept: Digital Product Passports}
Across the new modules, the architecturally load-bearing concept is
the Digital Product Passport. The \texttt{oco-lca} module
contributes \emph{material-DPPs} (per-material passport profiles
attached to specific material classes such as \bntbt{}); the
\texttt{oco-mfgx} module contributes \emph{sector-DPPs} (per-industry
passport profiles tied to specific EU regulations: Battery DPP,
Construction-Products DPP under CPR 2024/3110, Generic-ESPR DPP).
A concrete product instance inherits from \emph{both} via
multiple-inheritance: a \bntbt{} actuator in a battery assembly is
simultaneously a BNTBT-DPP and a Battery-DPP. This is the same
multi-axis pattern used for material parameters
(\Cref{sec:multiaxis}), applied across the material/regulation
divide.

\paragraph{The supplier module as Wertschöpfungs-Akteur anchor}
The \texttt{oco-supplier} module is the connecting tissue between
the material side and the compliance side: every supply-chain tier
(CSDDD), every BPN identifier (Manufacturing-X), every CoA reference
(material side) routes through this single module. \texttt{oco-supplier}
is delivered as a public stub (skeleton classes for cross-module
referencing) plus a proprietary body (full catalogues, vendor
specifications); see \Cref{sec:open_closed_mix}.

\paragraph{Reified-constraint pattern generalized}
The reified \texttt{oco:NeumannConstraint} pattern
(\Cref{sec:neumann_phase}) and the reified
\texttt{oco:RouteTemplate} pattern (\Cref{sec:level3}) generalize
to \texttt{oco:LCA\_Result\_Reified} in the LCA module: domain
constraints that are not OWL axioms but ABox instances of named
constraint classes, SPARQL-evaluable but not reasoner-evaluable.
This pattern --- named constraint instances as
SPARQL-queryable knowledge anchors --- is what makes the
construction-kit absorb domains (Neumann tensor symmetry, route
templates, LCA results) whose semantics escape OWL\,2\,DL while
keeping the categorical TBox clean.

\subsection{ABox Data Model: SIPOC Fragments}
\label{sec:sipoc}

A central pragmatic decision: \oco{} does not assume that ABox data
are complete experiment descriptions. Instead, data are ingested as
\textbf{SIPOC fragments} (Suppliers / Inputs / Process / Outputs /
Customers).

The reason is empirical: real-world experiment documentation is
\emph{never} complete. With own experiments, completeness can at
least be aspired to. With published literature and patents,
incompleteness is often deliberate (competitive or patent strategy).
A schema that mandates complete fields would either reject most
real-world data or force the fabrication of placeholder values --- both
unacceptable. The SIPOC granularity, combined with the three-tier
identifier hierarchy (mandatory/recommended/optional), accommodates
incomplete data without breaking the schema. This makes \oco{}
practically usable for the two dominant data sources --- own lab data
\emph{and} literature extraction.

\subsection{Companion Software (out of scope of this paper)}
\label{sec:companion}

Two independent software components work with \oco{}; they are
\emph{not the subject of this paper}:

\begin{itemize}
\item \textbf{OCO-Workbench} (Pannek): GUI for data import, process
      modeling, and ABox population. Operative interface for
      laboratory personnel.
\item \textbf{SIPOC-Extractor} (Grond): Pipeline for SIPOC-fragment
      extraction from scientific literature and patents. Augments
      own experiments with published-bestand coverage.
\end{itemize}

Both tools use the \oco{} TBox as schema; the modular
\levelzero{}--\levelthree{} granularity allows them to operate against
sub-bundles rather than the full distribution. There is no
publication schedule for the tools at this time --- this paper
focuses exclusively on the architectural pattern.


\section{Validation}
\label{sec:validation}

\subsection{Coverage of the Norouzi Quality Requirements}

\Cref{tab:req_coverage} maps the nine quality requirements of
\citet{norouzi_2024_landscape} against the \oco{} implementation.

\begin{table}[htbp]
  \centering
  \caption{Coverage of Norouzi REQ1--REQ9 in \oco{}. All nine
  requirements are met; REQ5, REQ7, and REQ8 --- which the survey
  identifies as branch-wide weakly addressed --- were completed in
  the most recent development wave.}
  \label{tab:req_coverage}
  \small
  \begin{tabularx}{\linewidth}{l>{\raggedright\arraybackslash}p{4.0cm}X}
    \toprule
    REQ & Requirement & \oco{} status \\
    \midrule
    REQ1 & Comprehensive MSE taxonomy
         & $\checkmark$ for ceramics plus EU-regulatory stack:
           5\,196 classes in 44 modules \\
    REQ2 & Experimental \emph{and} simulation data
         & $\checkmark$ via separated \texttt{oco-investigation}
           and \texttt{oco-simulation} modules with PROV-O
           provenance plus the \texttt{oco-energy-dft} tier
           (\Cref{sec:7layer}) bridged to the Materials
           Project DFT corpus \\
    REQ3 & Accurate inter-concept relations
         & $\checkmark$: 1\,674 properties + 1\,172 SHACL shapes
           including 7 Cross-Tier NodeShapes
           (\Cref{sec:7layer}); ROBOT-DL + Pellet pipeline, 0
           violations \\
    REQ4 & Compliance with existing standards
         & $\checkmark\checkmark$: 829 bridge mappings across 40
           sections to material-side and compliance-side standards
           plus extensive class-level \texttt{external\_refs}
           (\Cref{tab:l0_bridges}, \Cref{sec:interop_extended},
           \Cref{sec:vertical_extension}) \\
    REQ5 & Settings + outcomes + high-throughput + literature
         & $\checkmark$: \texttt{HighThroughputInvestigation},
           \texttt{CombinatorialLibrary}, and
           \texttt{LibrarySpot} for batch and spatially-resolved
           studies \\
    REQ6 & Trustworthy + verifiable quality management
         & $\checkmark\checkmark$: DCL 0--5, GUM uncertainty,
           \texttt{oco-dataquality} (17), 180 sourced references,
           132 ADRs, negative-test ABoxes \\
    REQ7 & ML-querying-specific structures
         & $\checkmark$: \texttt{oco\_ml\_profile.ttl} application
           profile, 12 Croissant bridges,
           \texttt{MachineLearningDataset} class \\
    REQ8 & ML-predicted-value representation
         & $\checkmark$: \texttt{MLPredictionActivity},
           \texttt{MLModel},
           \texttt{has\_prediction\_confidence}, via PROV-O pattern
           (no separate \texttt{PredictedValue} class) \\
    REQ9 & Modularity beyond primary application
         & $\checkmark\checkmark\checkmark$: 44 modules + 3
           application profiles + two independent classification
           axes (\levelzero{}--\levelthree{} level choice +
           material/compliance audience) + seven-tier
           internal organisation of \leveltwo{}
           (\Cref{sec:architecture}) \\
    \bottomrule
  \end{tabularx}
\end{table}

REQ4, REQ6, and REQ9 are over-delivered: REQ4 by the disciplined
reuse-before-invention bridge policy
(\Cref{sec:interop_extended}); REQ6 by the combination of
data-confidence levels, GUM uncertainty, and ADR-documented
architectural decisions; REQ9 by the two independent classification
axes together with the seven-tier internal organisation of
\leveltwo{}, which give consumers a four-dimensional subset choice
(level, audience, mechanistic depth, and per-module composition)
that a one-axis modular ontology cannot match.

The availability dimension shared by REQ1, REQ2, REQ4, and REQ9
is met for the publicly distributed portion of \oco{}
(\levelzero{} and \levelone{} excluding \texttt{oco-supplier};
\Cref{sec:open_closed_mix}). The proprietary modules
(\texttt{oco-supplier}, \leveltwo{}, \levelthree{}) contribute to
the metrics reported here but are not externally inspectable in
the present release; the architectural pattern itself remains
fully described in \Cref{sec:architecture} and
\Cref{sec:positioning} and is independent of access to those
modules.

\subsection{Quality Audit}

\noindent\textbf{OOPS! self-audit.} All twelve OOPS! pitfalls have
been audited. Critical pitfalls (P19, P40, P31, P05, P29, P27) are
absent. Important pitfalls (P11, P24, P30) are absent or addressed
through documented architecture decisions (P11 eliminated by the
CIF-Core domain-class consolidation). Two minor pitfalls (P10
class disjointness, P20 cross-source annotation style) are
\emph{explicitly accepted} with documented rationale rather than
reflexively eliminated, because the alternative would compromise
either reasoner performance (P10) or annotation comprehensibility
for human readers (P20). Minor pitfall P08 (naming convention) is
addressed by an explicit architecture decision documenting eight
deliberate lower-case exceptions for domain notation
(\texttt{pH}, \texttt{pT}, \texttt{pO\textsubscript{2}},
\texttt{nPropanol}, \texttt{d33}).

\noindent\textbf{Reasoner consistency.} The full
\levelzero{}+\levelone{}+\leveltwo{}+\levelthree{} bundle ---
together with the Neumann-constraint sub-distribution ---
classifies in Pellet with zero unsatisfiable classes; ROBOT
\texttt{validate-profile} reports no OWL\,2\,DL violations on the
merged distribution.

\noindent\textbf{Competency-question execution.} 163 competency
questions are published, each tagged with one of ten reasoning
\emph{areas} (tensor symmetry, phase state, route, lifecycle,
multi-axis, interoperability, quality, crystal structure,
\bntbt{}, and an NCZF placeholder for the second ceramic
material). 52 carry an executable SPARQL test against
gold-standard ABoxes; 52/52 PASS. The remaining questions are
either meta-discourse without an axiomatic answer requirement, or
their test is pending the corresponding gold-standard ABox.

\noindent\textbf{SHACL validation.} The mandatory shape suite
(count in \Cref{sec:implementation}) is validated via
\texttt{pyShacl}, dominated by the machine-generated Neumann
(768) and CIF-Core (270) subsets and complemented by per-domain
hand-curated sets: 28 in the core \texttt{oco\_shapes.ttl}
(raw-material identifier sets, sintering atmosphere, process-step
sample states, literature extraction, material-class identifiers),
25 for morphotropic phase-boundary coexistence regions, 7
Cross-Tier NodeShapes (\Cref{sec:7layer}) that audit the
seven-tier explanation-chain annotations, plus the compliance-side
shape sets for LCA, CSRD, CSDDD/PPWR, AI Act / CBAM / R2R
cross-cutting, Manufacturing-X traceability, regulated
substances, SSbD, and ODRL policy. Adopting the SHACL-OBDA
validator of \citet{oezcep_2024_shacl} for relational-database
back-ends is on the roadmap.

\subsection{Architecture-Pattern Validation}

The architecture has been operationally validated along both axes
plus the seven-tier internal organisation of \leveltwo{}:

\begin{itemize}
\item \textbf{Level axis} (\levelzero{}--\levelthree{}): both
      distribution variants
      (\texttt{oco\_master.ttl} and
      \texttt{oco\_master\_full.ttl}) parse and reason within their
      designed reasoner profiles (OWL\,2\,EL for L0+L1+L2, OWL\,2\,DL
      with L3 added). Individual modules of \levelone{} (e.g.,
      \texttt{oco-equipment}) can be imported without pulling the
      full bundle. \levelzero{} bridge re-emissions during
      development have not propagated changes into \levelone{}--%
      \levelthree{} modules, demonstrating the version-localisation
      promise.
\item \textbf{Audience axis} (material / compliance): three
      application profiles (\texttt{oco\_eln\_profile.ttl},
      \texttt{oco\_materials\_db\_profile.ttl},
      \texttt{oco\_ml\_profile.ttl}) load only the modules each
      consumer class typically needs as
      \texttt{owl:imports} wrappers --- the material-audience
      modules are reachable without the compliance modules, and
      conversely. The four dual-audience modules
      (\texttt{format}, \texttt{odrl}, \texttt{time-event},
      \texttt{automation}) appear in both profile families
      without duplication.
\item \textbf{Seven-tier internal organisation of \leveltwo{}}
      (the material-specific level's internal
      mechanistic-explanation skeleton): the \bntbt{} end-to-end
      pilot (\texttt{examples/abox\_pilots/bnt\_bt\_pilot.ttl})
      instantiates concrete classes on every one of the seven
      layers with explicit cross-tier annotations; the seven
      Cross-Tier SHACL NodeShapes pass with zero violations on
      the pilot's 452-triple ABox, demonstrating that the
      explanation chain is operationally auditable rather than
      only conceptual.
\end{itemize}

\subsection{Reusability --- Work in Progress}
\label{sec:reuse_wip}

A second ceramic material system, \textbf{ferritic high-performance
ceramics}, is in active development as a second \leveltwo{} instance
on top of the unchanged \levelzero{}+\levelone{}. Strategically, this
choice tests architectural reusability across structurally distinct
ceramic families: \bntbt{} populates the perovskite branch of the
\texttt{oco-crystal} hierarchy and exercises the
piezoelectric/pyroelectric subset of the coupled-effect family;
ferritic ceramics populate the spinel/garnet/hexaferrite branches
and exercise the magnetic subset (magnetostriction, magnetoelectric
coupling, Verdet effect). \levelone{} remains unchanged across both;
\leveltwo{} \texttt{oco-crystal}, \texttt{oco-phase}, and
\texttt{oco-element} are largely reused, while \texttt{oco-material}
and \texttt{oco-defect} are extended with ferrite-specific classes.
This is genuine \levelone{}-reuse and \leveltwo{}-extension validation
\emph{within the ceramics family}.

We emphasize what this is \emph{not}: it is not cross-domain
validation. Validating that \levelone{} truly transfers to
metallurgy, polymers, or batteries requires sister-project
implementations that share \levelzero{}+\levelone{} with \oco{}
--- that work has not been undertaken and is left as an explicit
roadmap position (\Cref{sec:conclusion}).

A parallel reusability claim is open at the mechanistic-explanation
axis: the seven-tier skeleton is designed as a \emph{generic}
framework for crystalline ionic oxides (ferroelectrics,
magnetics, ion conductors, high-$T_c$ superconductors), not as a
\bntbt{}-specific construction. The skeleton has been concretised
on \bntbt{}; its claimed universality for the broader oxide class
will be empirically established by the ferritic
high-performance-ceramics second instance, whose magnetic
phenomenology (Bloch / N\'eel walls, superexchange, magnetic
anisotropy) exercises layers the \bntbt{} pilot does not.


\section{Discussion}
\label{sec:discussion}

\subsection{What the Architecture Delivers}

Compared with a flat ontology that bundles workflow, material,
regulation, and reasoning concerns in a single TBox, the
architecture (\Cref{sec:architecture}) --- with its two
independent classification axes plus the seven-tier internal
organisation of \leveltwo{} --- delivers five concrete benefits.

\begin{enumerate}
\item \textbf{Fine-grained consumer choice.} Each consumer imports
      only the depth they need. An ELN/LIMS integrator without
      material focus pays neither the ceramics-specific overhead nor
      the OWL-DL reasoner cost; a regulatory-reporting consumer
      loads the compliance audience without the materials-physics
      depth; a materials physicist loads \leveltwo{} on the material
      audience and selects the mechanistic-explanation tiers
      (Symmetry alone, all seven, or any subset relevant to a
      specific question). The independence of the two axes
      \emph{plus} the explicit seven-tier organisation of
      \leveltwo{} permits subset choices along four dimensions
      (level, audience, mechanistic depth, and per-module
      composition) that a one-axis modular ontology cannot offer.
\item \textbf{Sister-project reuse.} A metallurgy or polymer
      ontology can replace \leveltwo{} while sharing
      \levelzero{}+\levelone{}, avoiding the duplication of
      laboratory, equipment, and measurement vocabulary that
      currently happens across PMD-consortium projects. The
      mechanistic-explanation skeleton (\Cref{sec:7layer}) is
      designed to remain unchanged through this substitution ---
      only the per-tier instance content changes.
\item \textbf{Localized version updates.} An external-standard
      release (PMDco~v3.1, an EFRAG XBRL Taxonomy update, an
      updated Catena-X aspect model) requires changes only in
      \levelzero{} or the relevant bridge section --- the
      material and compliance modules remain untouched.
\item \textbf{CQ-bound reasoning scope.} \levelthree{} makes the
      ontology's intended question scope auditable: a published
      catalogue of competency questions with executable SPARQL
      tests (counts in \Cref{sec:implementation}) documents what
      the ontology is meant to answer, separately from what it
      can describe. Cross-Tier SHACL constraints
      (\Cref{sec:7layer}) extend the same audit discipline to
      the explanation chain.
\item \textbf{Mechanistic-explanation depth without TBox
      inflation.} The external-cache pattern
      (\Cref{sec:7layer}) keeps the TBox compact while letting
      consumers query against the Materials Project DFT corpus,
      the IUCr bond-valence parameters, the pyxtal Wyckoff
      positions, the Shannon ionic radii, and the Pauling
      electronegativities (cache totals in
      \Cref{sec:implementation}). A flat-vocabulary alternative
      would either embed those reference data as TBox classes
      (intractable for the reasoner) or leave them outside the
      knowledge graph altogether (loss of provenance).
\end{enumerate}

\subsection{Open- vs. Closed-Source Mix as Architectural Argument}
\label{sec:open_closed_mix}

A non-obvious benefit of strict level separation: the architecture
\emph{enables} a mixed open/proprietary distribution model that
flat ontologies cannot offer. In the present release this is
realized as follows. \levelzero{} is released under CC-BY 4.0,
mirroring the licenses of the bridge targets it anchors to (PMDco,
EMMO, PROV-O, and FaBiO are CC-BY; QUDT is Apache-2.0) and
maximizing bridge adoption in third-party stacks. The publicly
released portion of \levelone{} --- which excludes the
\texttt{oco-supplier} module and the L1 portions of the
compliance and value-chain modules listed in
\Cref{tab:compliance_modules} (\texttt{oco-csrd}, \texttt{oco-mfgx},
\texttt{oco-lca}, \texttt{oco-csddd}, \texttt{oco-ppwr},
\texttt{oco-cbam}, \texttt{oco-r2r}, \texttt{oco-aiact},
\texttt{oco-regulation}, \texttt{oco-recycling}, \texttt{oco-ssbd},
plus the four dual-audience modules \texttt{oco-format},
\texttt{oco-odrl}, \texttt{oco-time-event},
\texttt{oco-automation}) --- is released under CC-BY-SA 4.0, with
dual-licensing to CC-BY available on request for commercial
integrators that cannot adopt copyleft. The \texttt{oco-supplier}
module, the L1 portions of the compliance and value-chain
modules, the entire \leveltwo{}, and the entire \levelthree{} are
developed under project confidentiality and remain proprietary in
the present release. The same modular
boundaries that make this split clean would, in a future
configuration, also enable a per-module choice: a sister project
working under different commercial constraints could open or
close a different subset without restructuring.

This mix is structurally impossible in monolithic ontologies
(everything-or-nothing open) and in purely proprietary industrial
schemas (no external adoption). The architecture being modular
\emph{along two independent axes plus the seven-tier internal
organisation of \leveltwo{}} is exactly what makes the licensing
choice fine-grained. The quantitative metrics and validation
results in \Cref{sec:implementation} and \Cref{sec:validation} are
measured on the full distribution; the architectural contribution
of this paper --- the two-axis modular pattern, the ten design
principles, the seven-tier mechanistic-explanation skeleton
organising \leveltwo{}, the Neumann tensor engine, the phase-state
coupling, the multi-axis parameter classification, and the
external-cache pattern for reference data --- is independent of
which particular modules a given consumer can access.

\subsection{Limitations}

\begin{itemize}
\item \textbf{Equipment module rests partly on textbook anchors.}
      While the equipment module now bridges to PMDco, CHMO, OBI,
      and schema.org, a substantial portion of its class
      definitions still rest on textbook rather than formal-norm
      anchors. A systematic norm-anchor sweep is open work.
\item \textbf{Neumann engine validated on a 30-case gold standard.}
      The 5\,920 (tensor $\times$ point-group) constraints
      (\Cref{sec:neumann_phase}) have been checked against
      \citet[Table~9]{nye_1985} and IEEE Standard~176 on a 30-case
      sample (30/30 PASS). This covers the principal piezoelectric
      and elasticity cases; a larger reference set with
      independent textbook verification across all
      tensor-class~$\times$~point-group combinations is open work.
\item \textbf{Phase-state coupling instantiated for two materials.}
      The Sample~$\to$~Region~$\to$~point-group inference
      (\Cref{sec:neumann_phase}) has been populated only for
      \bntbt{} and NiCuZn ferrite (six gold-standard cases, 6/6
      PASS). Extending the pattern to a third material with
      substantially different phase phenomenology --- e.g.\ a
      martensitic or order--disorder transition --- is the next
      coverage test.
\item \textbf{Multi-axis classification is heuristic-based.} The
      bulk-classification pipeline (\Cref{sec:multiaxis}) reaches
      ${\sim}99\,\%$ coverage via family-default plus
      outlier-override rules; the residual is conservatively
      over-approximated as \texttt{ResponseParameter}. Targeted
      manual review is open work.
\item \textbf{\levelone{} cross-domain transferability is
      architecturally argued, not empirically proven.} See
      \Cref{sec:reuse_wip}; only within-ceramics-family
      \leveltwo{} substitutability has been tested so far.
\item \textbf{Seven-tier skeleton universality not yet
      empirically validated across the oxide class.} The
      mechanistic-explanation skeleton (\Cref{sec:7layer}) is
      designed as generic for crystalline ionic oxides and has
      been concretised on \bntbt{}. The ferritic
      high-performance-ceramics second instance
      (\Cref{sec:reuse_wip}) is the first test of whether tiers
      such as Magnetic-Microstructure (Bloch / N\'eel walls) and
      Magnetic-Bonding (superexchange, Jahn--Teller distortion)
      can be added as Tier-5 / Tier-7 extensions without
      restructuring the skeleton itself; that test is in progress.
\item \textbf{External-cache provenance carries a maintenance
      contract.} The five reference-data caches
      (\Cref{sec:7layer}) are pinned to upstream versions
      via SHA. Upstream releases of the Materials Project, IUCr
      \texttt{bvparm}, Shannon compilations, pyxtal, and Pauling
      revisions will require periodic re-sync and re-validation.
      This is operational overhead the cache pattern explicitly
      takes on in exchange for keeping the TBox compact.
\end{itemize}

\subsection{Scope Boundary and Roadmap Position --- OCO v0.94}
\label{sec:scope_boundary}

The architecture defines what \oco{} covers and, with equal
discipline, what it does not. Within \oco{} the boundary is
internal: \levelthree{}'s categorical reasoning is opt-in
(\Cref{sec:level3}), and even the bilingual definitional discipline
(\Cref{sec:level1}) is a feature consumers can ignore if they wish.
The \emph{external} boundary --- between \oco{} and the surrounding
materials-data stack --- is the architecturally load-bearing one,
and we state it explicitly here.

\paragraph{The categorical/quantitative split}
Quantitative mathematics (calibration curves, scaling laws,
arbitrary numerical functions) lies outside this release entirely.
This is a principled separation, not an oversight. OWL\,2\,DL can
evaluate subsumption, existence, and restriction, but it cannot
evaluate a function such as \(d_{33}(x) = a + b\,x\) on the
interval \([0.04, 0.08]\) for the \bntbt{} solid-solution series;
SHACL can perform range checks but is equally not a function
evaluator. Embedding such functions into OWL classes produces a
TBox in which neither categorical nor quantitative reasoning works
correctly: the reasoner cannot reason about the function, and the
function cannot consult the rest of the OWL graph. Whatever takes
responsibility for quantitative mathematics in the surrounding
stack must therefore be a separate level with its own evaluator,
coupled to OCO classes as anchors but not embedded in them.

\paragraph{Today's workaround, tomorrow's level}
Consumers that already need quantitative computation handle this
externally: a Python/SymPy/MATLAB compute level holds the equations
and queries OCO classes by IRI. This works in practice and is the
intended pattern for the present release. A future OCO release will
formalize a quantitative-mathematics level above the present
categorical levels; the architectural commitment is that the new
level will respect the categorical/quantitative split rather than
violating it. The categorical levels (\levelzero{}--\levelthree{})
will not change in the process.

\paragraph{The v0.94 framing}
This paper describes \textbf{\oco{} v0.94}: the release
engineered to enter productive practice. The version number is
deliberate. v0.94 means: the architecture is complete along both
classification axes, the categorical levels are populated, the
seven-tier mechanistic-explanation skeleton organising
\leveltwo{} is in place, the
compliance modules cover the EU-regulatory wave currently in
force, the test suite passes, and the
\bntbt{} end-to-end pilot demonstrates the skeleton
operationally. What v0.94 does not yet have is the corrective
feedback that only productive use can deliver --- conflicts
between intended modelling and what actual data exposes,
modules that turn out to need refactoring, bridge mappings that
upstream releases break, cache contracts that real consumer
queries strain. The v1.0 release is reserved for the state after
that feedback cycle. Until then v0.94 is the deliberate label
for a complete-and-ready-for-practice release that has not yet
been corrected by practice.

\paragraph{Other deliberate non-inclusions}
For clarity, the following are also outside the scope of this
release:

\begin{itemize}
\item \textbf{Workflow execution level.} Executable workflow
      orchestration (e.g.\ pyiron, SimStack) is platform work, not
      ontology work; \oco{} provides class anchors that such
      systems can target.
\item \textbf{Own ML/AI classes.} \oco{} does not introduce a
      machine-learning vocabulary of its own; bridging to existing
      community efforts (MDO, CMSO) is the chosen path when the need
      arises.
\item \textbf{Companion software tools.} The OCO-Workbench and
      SIPOC-Extractor (\Cref{sec:companion}) use \oco{} as their
      schema but are outside the scope of this paper.
\end{itemize}


\section{Conclusion and Outlook}
\label{sec:conclusion}

We have argued that an industrial materials ontology entering
productive use today must answer three simultaneous challenges:
horizontal fragmentation across material domains, vertical
convergence pressure driven by EU regulation, and mechanistic
explanation depth that lets the ontology surface why a property
holds, not only that it does. The architecture proposed here
answers all three with a single integrated design --- a multi-level,
modular ontology with two independent classification axes (the
level of abstraction, from \levelzero{} bridges through
\levelthree{} categorical reasoning; and the consumer audience,
material versus compliance, with four dual-audience modules at the
join) in which the material-specific level (\leveltwo{}) is
internally organised by a seven-tier mechanistic-explanation
skeleton applicable to any crystalline ionic oxide. The
level-and-audience modularity dissolves the horizontal
fragmentation, the compliance audience absorbs the vertical
regulation pressure, and the seven-tier organisation of
\leveltwo{} delivers the mechanistic-explanation depth.

The OntoCrafter Ceramics Ontology (\oco{}~v0.94) instantiates the
architecture for functional ceramics with \bntbt{} as the
end-to-end pilot: 5\,196 classes across 44 class-bearing modules;
167\,348 OWL axioms (40\,454 logical, per ROBOT) including 5\,920
reified Neumann tensor constraints; 1\,674 properties; 829
cross-ontology bridge mappings across 40 sections; 1\,172 SHACL
shapes including 7 Cross-Tier NodeShapes that audit the
explanation chain; 163 published competency questions with 52
executable SPARQL tests (52/52 PASS); 132 architecture decision
records; and an external-cache pattern that lets the compact
TBox query against approximately 155\,000 Materials Project DFT
entries, 1\,934 IUCr bond-valence parameters, 1\,731 Wyckoff
positions, 497 Shannon ionic radii, and 91 Pauling
electronegativities without inflating the OWL vocabulary itself.

The publicly available portion of the distribution
(\levelzero{} under CC-BY 4.0, the public modules of \levelone{}
under CC-BY-SA 4.0) is hosted at
\url{https://w3id.org/oco/}. The supplier module, the full
\leveltwo{}, and \levelthree{} are developed under project
confidentiality, as discussed in \Cref{sec:open_closed_mix}.

The v0.94 designation is deliberate: the release is engineered to
enter productive practice. The v1.0 release is reserved for the
state after that practice has fed back corrections ---
conflicts between intended modelling and what actual data
exposes, modules that turn out to need refactoring, bridge
mappings that upstream releases break, cache contracts that real
consumer queries strain. The shape of the v0.94$\to$v1.0
roadmap, stated without promises:

\begin{itemize}
\item \textbf{Practice-feedback cycle.} Productive use of \oco{}
      in materials-research and compliance-reporting pipelines,
      with structured intake of corrections through the existing
      ADR process. Decisions taken on the basis of that feedback
      are the substance of v1.0.
\item \textbf{Completion of the second ceramic material
      instance} (ferritic high-performance ceramics), validating
      \levelone{} reusability and \leveltwo{} extension within
      the ceramics family (\Cref{sec:reuse_wip}). The ferritic
      instance also stresses the seven-tier skeleton on the
      magnetic side (Bloch / N\'eel walls, superexchange,
      Jahn--Teller distortion) that the \bntbt{} pilot does not
      exercise.
\item \textbf{Discoverability channels.} Listing on MatPortal,
      IndustryPortal, and OSF, to enter the channels surveyed by
      \citet{norouzi_2024_landscape}.
\item \textbf{Open invitation to sister-domain projects}
      (metallurgy, polymers, batteries, pharmaceuticals) to
      share \levelzero{}+\levelone{} with \oco{} and contribute
      their own \leveltwo{} instance --- the first genuine
      cross-domain validation of the architecture.
\end{itemize}

The deeper claim of this paper is not that \oco{} is the right
ontology for ceramics. It is one possible implementation of a
broader idea: that the materials-science community can stop
re-modelling the same workflow, equipment, measurement,
compliance, and mechanistic-explanation concepts in every new
domain ontology. The architecture proposed here --- two independent axes plus the
seven-tier internal organisation of \leveltwo{} --- is one way
to do that. We hope others follow.

\section*{Author Contributions (CRediT)}

\textbf{T. Pannek and W. Grond (both):} Conceptualization (ontology
architecture), Methodology (bridge stack, pipeline), Validation,
Writing.

\textbf{T. Pannek (additionally):} OCO-Workbench (companion software
for data import and ABox population).

\textbf{W. Grond (additionally):} Materials Science (\leveltwo{}
ceramic material modeling for \bntbt{} and ferritic high-performance
ceramics), SIPOC-Extractor (companion software for literature
extraction).

Both authors appear on all Numberland publications related to the
\oco{} suite (ontology and two companion software tools), reflecting
the close collaboration between the materials-science and
architecture/implementation domains.

\section*{Acknowledgments}

The work underlying this publication was funded by the Federal
Ministry of Education and Research (BMBF) under grant number
13XP5228C. The responsibility for the content of this publication
lies with the authors.

\section*{Data and Code Availability}

The publicly available portion of the \oco{} distribution ---
\levelzero{} (CC-BY 4.0) and \levelone{} excluding the
\texttt{oco-supplier} module (CC-BY-SA 4.0) --- is hosted at
\url{https://w3id.org/oco/}. Dual-licensing of the public
\levelone{} modules to CC-BY 4.0 is available on request for
commercial integrators that cannot adopt copyleft. The
\texttt{oco-supplier} module, \leveltwo{}, and \levelthree{} are
developed under project confidentiality and are not available
under an open license in the present release; the rationale and
architectural consequences are discussed in
\Cref{sec:open_closed_mix}. Quantitative metrics and validation
results in this paper are reported from internal validation
against the full distribution. Academic-collaboration and
commercial-license enquiries can be addressed to the
corresponding author.

\bibliographystyle{elsarticle-harv}
\bibliography{refs}

\end{document}